\documentclass[12pt]{iopart}
\usepackage{url}
\usepackage{cite} \usepackage{hyperref} \usepackage{graphicx}
\usepackage{amsfonts} \usepackage[latin1]{inputenc}
\usepackage[english]{babel}

\usepackage{todonotes}
  \def\nuc#1#2{\relax\ifmmode{}^{#1}{\protect\text{#2}}\else${}^{#1}$#2\fi}
  \def\itnuc#1#2{\setbox\@tempboxa=\hbox{\scriptsize\it #1}
    \def\@tempa{{}^{\box\@tempboxa}\!\protect\text{\it #2}}\relax
    \ifmmode \@tempa \else $\@tempa$\fi}
  \newcommand{\nm}{\ensuremath{N_\mathrm{max}}}
  \newcommand{\ho}{\ensuremath{\hbar \Omega}}
  \newcommand{\eqref}[1]{(\ref{#1})}

\begin{document}

\title{Living on the edge of stability, the limits of the nuclear
  landscape}

\author{C.~Forss\'en} \address{Department of Fundamental Physics,
  Chalmers University of Technology, SE-412 96 G\"oteborg, Sweden}
\author{G.~Hagen} \address{Physics Division, Oak Ridge National
  Laboratory, Oak Ridge, TN 37831, USA} \address{Department of Physics
  and Astronomy, University of Tennessee, Knoxville, TN 37996, USA}
\author{M.~Hjorth-Jensen} \address{National Superconducting Cyclotron
  Laboratory and Department of Physics and Astronomy, Michigan State
  University, East Lansing, MI 48824, USA} \address{Department of
  Physics and Center of Mathematics for Applications, University of
  Oslo, N-0316 Oslo, Norway} \author{W.~Nazarewicz}
\address{Department of Physics and Astronomy, University of Tennessee,
  Knoxville, TN 37996, USA} \address{Physics Division, Oak Ridge
  National Laboratory, Oak Ridge, TN 37831, USA} \address{Institute of
  Theoretical Physics, Warsaw University, PL-00681 Warsaw, Poland}
\author{J.~Rotureau} \address{Department of Fundamental Physics,
  Chalmers University of Technology, SE-412 96 G\"oteborg, Sweden}

\begin{abstract}
  A first-principles description of nuclear systems along the drip
  lines presents a substantial theoretical and computational
  challenge.  In this paper, we discuss the nuclear theory roadmap,
  some of the key theoretical approaches, and present selected results
  with a focus on long isotopic chains.  An important conclusion,
  which consistently emerges from these theoretical analyses, is that
  three-nucleon forces are crucial for both global nuclear properties
  and detailed nuclear structure, and that many-body correlations due
  to the coupling to the particle continuum are essential as one
  approaches particle drip lines. In the quest for a comprehensive
  nuclear theory, high performance computing plays a key role.
\end{abstract}

 \maketitle

\section{Introduction}
\label{sec:introduction}
To understand why nucleonic matter is stable is one of the overarching
aims and intellectual challenges of basic research in nuclear
physics. To relate the existence and properties of nuclei to the
underlying fundamental forces and degrees of freedom, is central to
present and planned rare isotope facilities, see for example
Refs.~\cite{RISAC,Decadal2012,jacekwitek1998,thoennessen2011,brown2001,michael2012,sorlin2008,witeknature2012}.
Important properties of nuclear systems are binding energies, radii,
density distributions of nucleons, spectra, and decays.  These
quantities convey important information on the individual-nucleon
motion manifesting itself in the shell structure of nuclei, including
the appearance and disappearance of magic numbers, interplay between
high-$j$ unique-parity orbits and natural-parity states, and the rapid
changes of nuclear properties around the reaction thresholds.

To relate the stability of nucleonic matter at various energies and
length scales\footnote[1]{A simultaneous description of nuclei and
  neutrons stars in terms of hadronic degrees of freedom constitutes a
  daunting theoretical challenge as the length scale spans over 19
  orders of magnitude, from several $10^{-15}$\,m (nuclear radii) to
  approximately 12 kilometers (neutron star radii).} to the underlying
fundamental forces is a major quest for theoretical modeling.  A
multiscale approach is required which incorporates different degrees
of freedom at relevant length and energy scales. Unfortunately,
quantum chromodynamics (QCD), the underlying theory of the strong
interaction, is highly non-perturbative in the energy region
characteristic of nuclear structure. This region is governed by
nucleonic (and sometimes mesonic) degrees of freedom. This requires an
effective theory, usually framed as  chiral effective field theory
(EFT), which is consistent with the symmetries of low-energy QCD and
the separation of scales relevant to the low-energy nuclear many-body
problem.  Linking different scales is far from easy.  A key challenge
is to understand the link between QCD and effective field theories.
Interactions derived from these low-energy theories carry also a
dependence on an energy scale, defined in terms of an energy cutoff
$\Lambda$, that separates the Hilbert space of interest from its
higher-energy complement. The cutoff is usually chosen so that it is
possible to reproduce nucleon-nucleon (NN) scattering phase shifts up
to $\sim 300$~MeV laboratory kinetic energy, which requires
$\Lambda\sim 500-700$ MeV. For details, we refer the reader to the
recent reviews~\cite{epelbaum2009,machleidt2011}.

An important question is to understand the role of many-nucleon forces
from chiral EFT in the nuclear medium.  Furthermore, using EFT-based
interactions in a many-body environment entails the development of
proper many-body theories that allow for first principle
calculations. There are other issues as well, such as the estimation
of theoretical errors.  For example, most many-body methods applied to
the nuclear many-body problem involve basis set expansions.  The
errors which arise due to basis truncations need proper
quantifications and clarifications \cite{kvaal2009,HOerrors,hjorthjensen2010,coon2012}.  
Finally, a proper
link between first-principle methods (which are of limited use when
very many degrees of freedom are at play) and approaches based on the
density functional theory, is essential if one wishes to understand
nuclei and nuclear matter from a bottom-up perspective
\cite{UNEDF,FurNPN}.

\begin{figure}[thbp]
  \begin{center}
  \includegraphics[width=0.6\textwidth,clip=]{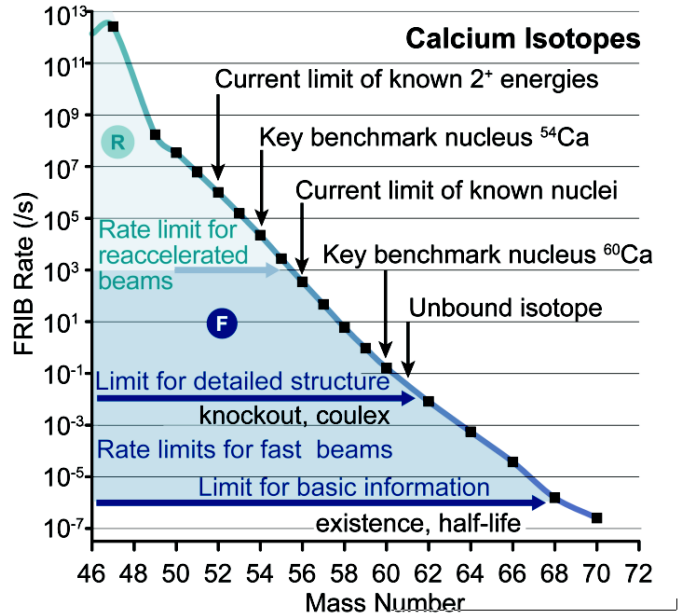}
  \end{center}
  \caption{Expected experimental information on the calcium isotopes
    that can be obtained at FRIB. The limits for detailed
    spectroscopic information are around $A\sim 60$. (Courtesy of Brad
    Sherill \cite{bradprivatecomm2012}.)}
\label{fig:calciumreach}
\end{figure}

Neutron-rich nuclei are particularly interesting for this endeavor. As
the neutron number increases within a particular chain of isotopes,
one eventually reaches the limit of stability, the neutron drip line,
where isotopes with additional neutrons are not neutron-bound
anymore. The appearance or absence of magic numbers, formation of
neutron skins and halos, and detailed tests of shell structure at the
limit of neutron-to-proton asymmetry can be probed via investigations
of masses, radii, and excited states of neutron-rich nuclei. Examples
of recent progress in the radioactive nuclear beam (RNB) science are
measurements of masses with Penning traps coupled to radioactive beams
and nuclear charge radii and moments using laser spectroscopy
techniques. New experimental data in the neutron-rich territory are
crucial for constraining theoretical models of nuclei and
astrophysical processes, see, e.g.,
Refs.~\cite{schatz1998,langanke2001}. To put things in perspective,
Fig.~\ref{fig:calciumreach} shows the expected experimental
information on the Ca isotopic chain that will be obtained at the
future Facility for Rare Isotope Beams (FRIB).

Having access to precise measurements of masses, radii, and
electromagnetic moments for a wide range of nuclei allows to study
trends with varying neutron excess. A quantitative description of
various experimental data with quantified uncertainty still remains a
major challenge for nuclear structure theory.  Global theoretical
studies of isotopic chains, such as the Ca chain in
Fig.~\ref{fig:calciumreach}, make it possible to test systematic
properties of effective nuclear interactions. Such calculations also
provide critical tests of limitations of many-body methods. As one
approaches the particle emission thresholds, it becomes increasingly
important to describe correctly the coupling to the continuum of
decays and scattering channels \cite{michel2009,NazPPNP}. While the
full treatment of antisymmetrization and short-range correlations has
become routine in {\em ab initio} approaches to nuclear bound states, the
many-body problem becomes more difficult when long-range correlations
and continuum effects are considered.

This paper is organized as follows. Section \ref{sec:formalism}
outlines modern theoretical approaches to the nuclear many-body
problem, starting with renormalization schemes for nuclear forces and
ending with a discussion of various many-body methods, in particular
aimed at describing systems near the nucleon drip lines.  Section
\ref{sec:results} presents selected results for stable and weakly
bound systems, with an emphasis on long isotopic chains.  Conclusions
and suggestions for future work are contained in
Sec. \ref{sec:conclusions}.

\section{Theoretical foundations}\label{sec:formalism}
The task of developing a comprehensive theoretical framework that
would be quantitative (i.e., capable of reproducing and explaining
existing experimental data), have predictive power (i.e., capable of
massive extrapolations), and provide uncertainty quantification (i.e.,
theoretical error bars) is daunting.  To this end, theoretical models
must meet three stringent requirements: (i) input (interactions,
functionals) must be of high quality; (ii) many-body dynamics and
correlations must be accounted for; and (iii) the associated formalism
must take care of open-quantum-system aspects of the nucleus.

Solving the nuclear many-body problem is a challenging endeavor, since
``exact'' solutions exists only for the very lightest systems and
closed form solutions only for highly-idealized, non-realistic
cases. Over the last few decades there has been tremendous progress in
developing many-body {\em ab initio} techniques capable of treating light
and medium-mass systems.  The state-of-the-art methods are based on
controlled approximations and the underlying computational schemes
account for successive many-body corrections in a systematic way.

The currently used theoretical techniques include: coupled-cluster
methods \cite{shavittbartlett2009,helgaker,hagen2010a}, Quantum Monte
Carlo applications \cite{pieper2001a,kdl1997,utsuno1999}, perturbative
expansions \cite{hjorthjensen1995}, Green's function methods
\cite{dickhoff,barbieri2004}, correlation operator methods
\cite{roth2005}, the density-matrix renormalization group
\cite{white1992,schollwock2005,rotureau2009}, density
functional theory \cite{witeknature2012,bartlett2005}, in-medium similarity
renormalization group \cite{tsukiyama2011,hergert2012}, and large-scale
diagonalization methods \cite{caurier2005,navratil2009}.

In low-energy nuclear physics, baryons (protons, neutrons, and
possibly deltas), and mesons (pions and possibly other mesons with
masses below 1 GeV) are considered to be the appropriate degrees of
freedom, meaning that one can derive effective Hamiltonians based on
the interactions between these constituents. Those Hamiltonians are
derived from Lagrangians that are consistent with the symmetries of
low-energy QCD using chiral perturbation theory at different orders
($\nu$), in terms of a perturbative expansion using a hard scale
$\Lambda$ and a soft scale $Q$ \cite{machleidt2011}.  Within this
picture, three-nucleon forces (3NFs) arise naturally since the
interactions are derived for effective degrees of freedom at low
energy, and heavier baryons and mesons are integrated out.

Currently, there are at least three different many-body methods for
nuclear structure with the capability to include both nucleon-nucleon
interactions and 3NFs, yielding results that meet the few-body
benchmarks. For light nuclei (with mass numbers $A \sim 12$), the
Green's function Monte Carlo
\cite{pudliner1997,pieper2001,wiringa2002} and large-basis no-core
shell-model (NCSM) approaches
\cite{navratil1998,navratil2000b,navratil2009} have been successfully
applied.  These methods provide excellent results for two- and
three-body Hamiltonians applied to light nuclei.  However, for
medium-mass and heavier nuclei, the dimensionality of the
many-particle problem becomes intractable for these techniques.
More recently, the coupled-cluster method has been applied to the
structure of light and medium-heavy nuclei
\cite{hagen2010a,hagen2012a,roth2012,binder2012}, providing excellent
benchmarks for few-body systems such as $^4$He \cite{hagen2010a}. This
technique is best suited for treating nuclei around closed-shells but
has very advantageous scaling properties that enable accurate
calculations in very large model spaces. At the singles and doubles
level of the method (expanding Slater determinants in terms of the
exponential of one-particle-one-hole and two-particle-two-hole
correlations), the number of floating-point operation scales as
$n_o^2n_u^4$, where $n_0$ and $n_u$ are the numbers of hole and
particle orbitals, respectively.  Such soft scaling, when compared to
the nearly combinatorial scaling of methods based on Hamiltonian
diagonalization (as a function of basis size and/or particle number),
allows one to build an extension of {\em ab initio} descriptions of nuclei
all the way to medium and heavy systems.

Nuclear {\em ab initio} methods are now evolving to tackle a crucial
challenge; the description of open nuclear systems. This development
is timely since the overarching scientific questions of modern nuclear
structure are about very neutron-rich or proton-rich nuclei whose
properties are impacted by a coupling to the particle continuum of
scattering and decaying states.  Many nuclei of interest lie every
close to particle emission thresholds, i.e., they are either
short-lived or unstable. Such nuclei cannot be described in a closed
quantum formulation, which assumes that the nucleons are artificially
confined in a trap; hence, they are unable to decay.

It is therefore crucial for the nuclear many-body problem that new
theoretical methods are developed that allow for an accurate
description of loosely bound and unbound nuclear states.  The recently
developed complex-energy Gamow Shell Model (GSM)\cite{GSMrev} has
proven to be a reliable tool in the description of nuclei, where
continuum effects cannot be neglected. In the GSM, a many-body basis
is constructed from a single-particle Berggren ensemble
\cite{berggren1968,Beggren1993,lind1993}, which treats bound,
resonant, and scattering states on equal footing. Recently, GSM
calculations of loosely bound and unbound states in nuclei, starting
from a realistic interaction and a Gamow-Hartree-Fock basis were
reported~\cite{hagen2006}. However, an {\em ab initio} description of light,
unstable nuclei within the GSM approach will require novel many-body
truncation schemes. In Ref.~\cite{hagen2007d} the Berggren basis was
employed for the first time in coupled-cluster calculations of the
helium isotopes, and there are promising attempts to include the
Berggren basis in an NCSM-like framework~\cite{papadimitriou-unpub}.

Finally, for the heavy and complex nuclei, the tool of choice is the
nuclear Density Functional Theory (DFT) \cite{bender2003} and its
extensions. Modern nuclear DFT is based on the mean-field approach
rooted in the self-consistent Hartree-Fock-Bogoliubov (HFB) problem.
The DFT work is closely tied to {\em ab initio} studies of experimentally
inaccessible systems such as neutron drops to enhance the predictive
capabilities \cite{GandolfiDrops,BognerDrops,UNEDF1}. Important areas
of research are the structure and decays of very neutron-rich nuclei,
the dynamics of the fission process in heavy nuclei, and the structure
of neutron star crust.

The quasiparticle HFB energy spectrum contains discrete bound states,
resonances, and nonresonant continuum states
\cite{dobaczewski1984,dobaczewski1996,HFBPairing}. That is, the effect
of the particle continuum is naturally incorporated into the
formalism, provided that the system of interest is particle-bound.  In
order to treat weakly-bound nuclei accurately, special care should be
paid to the spatial extension of HFB states \cite{HFBPairing}. In this
context, of particular interest are coordinate-space HFB approaches in
large boxes \cite{Pei2011}, and PTG-HFB \cite{PTGHFB} and Gamow-HFB
\cite{GamowHFB} frameworks.

The effective interaction in DFT, represented by the energy density
functional (EDF), is characterized by a set of coupling constants
constrained by experimental data and pseudo-observables determined by
{\em ab initio} calculations.  The uncertainty margins on EDF parameters are
obtained by means of statistical methods like linear-regression with
error analysis, which allows us to determine the correlations among
EDF parameters, parameter uncertainties, and the errors of calculated
observables
\cite{Klupfel,reinhard2010,kortelainen2010,UNEDF1,FatPie11}. Such an
approach is essential for providing predictive capability and
extrapolability, and for estimating the theoretical uncertainties. A
representative example of DFT calculations containing uncertainty
quantification, highly relevant to this paper, are the large-scale DFT
calculations of Ref.~\cite{witeknature2012} assessing the limits of
the nuclear landscape. Quantifying the limits of nuclear binding is
important not only for understanding the mechanism of nuclear binding,
but also for understanding the origin of elements in the
universe. Indeed, the astrophysical processes responsible for the
generation of many heavy elements operate very closely to the drip
lines; hence, the structure of very exotic, weakly bound nuclei
directly impacts the way the elements are produced in stars.


In the following, we shall first briefly discuss models for the
nuclear interactions and how these can be renormalized for truncated
Hilbert spaces. Thereafter follows a brief description of
configuration-interaction and coupled-cluster approaches to the
nuclear many-body problem. We then discuss some developments that
allow the study of open quantum systems within {\em ab initio} frameworks.

\subsection{Realistic nuclear interactions}

Ideally, the nuclear interactions should be derived from QCD, the
underlying theory for the strong interaction. Unfortunately, the
$^1S_0$ NN scattering length, a fundamental quantity characterizing
the NN interaction, still remains a huge challenge for  lattice QCD
(LQCD) calculations, as the physical point (at the actual pion mass of
140 MeV) lies in a resonance in the unitary region \cite{Beane06}.  In
order to improve calculations, one needs to go to smaller lattice
spacings, around $b=0.05$\,fm, but such computations require lattices
of the order of $96^3\times 193$ (the simulation cost grows as
$b^{-6}$). Meanwhile, the LQCD computations for NN systems
\cite{ishii2007,ishii2010,Beane12,Beane12a} provide important insights
about the basic properties of the nuclear force.

Several models of the NN interaction have been developed during recent
years. These interactions reproduce NN scattering data up to 300-350
MeV laboratory energy with excellent precision
\cite{machleidt2001,wiringa1995,nijmegen1994}.  The number of free
parameters used in the fitting procedure is around 40.  Being
optimized to on-shell scattering data, various interaction models
introduce different off-shell behavior resulting in different
predictions for finite nuclei.  (For more discussion on off-shell
effects, see, e.g., Ref.~\cite{engvik1997}.)  The interaction models
of Refs.~\cite{machleidt2001,wiringa1995,nijmegen1994} are either
based on one-boson exchange with selected low-mass mesons, or are
simply expressed in terms of local operators and optimized to data.
More recently, due to a progress made in chiral-EFT
\cite{weinberg1990,vankolck1999}, several groups have constructed
interaction models based on the underlying symmetries of QCD.

The starting point of the chiral-EFT is an approximation to the QCD
Lagrangian. The chiral effective Lagrangian is given by an infinite
series of terms with an increasing number of derivatives and/or
nucleon fields, with each term being dependent on the pion field
according to the rules of the broken chiral symmetry.  Applying this
Lagrangian to the NN scattering results in a systematic series of
Feynman diagrams in a small parameter, the ratio of $Q/\Lambda_\chi$,
where $Q$ is pion mass and $\Lambda_\chi \approx 1$ GeV is the chiral
symmetry breaking scale.  For a given order $\nu$, the number of
contributing terms is finite and calculable; these terms are uniquely
defined, and the prediction at each order is model independent.  By
going to higher orders, the amplitude can be calculated to any desired
accuracy.  This scheme has become known as chiral perturbation theory
($\chi$PT).

Interactions derived from EFT have a number of advantages over
traditional potential models. First, the corresponding currents are
consistently formulated, and this is important for the correct
description of observables other than the energy. Second, a power
counting in terms of $Q/\Lambda_\chi$ exists for systematic
improvements of the interaction and observables. Third, the hierarchy
of NN forces, 3NFs, and higher-rank forces is explained by power
counting. Three-nucleon forces enter at next-to-next-to-leading order
(N$^2$LO) order, and four-nucleon forces appear at N$^3$LO.  Recent
work \cite{machleidt2011,epelbaum2011} provides, for the first time, a
chiral interaction of quantitative accuracy.  The authors of
Ref.~\cite{entem2003} undertook the task of generating an accurate NN
interaction based on chiral perturbation theory at N$^3$LO order.  The
number of free parameters used in this chiral interaction is 24, which
is similar to the number of free parameters used to parametrize other
NN forces.  Three-nucleon forces have been explored recently in light
nuclei and neutron and nuclear matter
\cite{kalantar2012,hebeler2011}. In light nuclei, chiral 3NFs affect
the binding energy, radii and transition
probabilities~\cite{navratil2009}. They are responsible for the
anomalous long half life of $^{14}$C~\cite{maris2011}. In oxygen
isotopes, they are believed to determine the position of the neutron
drip line and the structure of neutron-rich
isotopes~\cite{otsuka2010,hagen2012a}. In calcium isotopes, 3NFs are
important for our understanding of shell evolution and (sub)shell
closures~\cite{holt2012,hagen2012b}. Three-nucleon forces also affect
properties of neutron matter~\cite{hebeler2010,GandolfiNS,steiner12}
and the saturation of nuclear matter~\cite{hebeler2011}.

\subsection{Derivation of effective interactions for truncated Hilbert spaces}
\label{sec:effective}

The nucleon-nucleon interaction is strongly repulsive at short
distances, meaning that a calculation which starts with the free
interaction may converge very slowly.  To overcome this problem, one defines effective interactions
which can be used as starting points for calculations with
basis-expansion approaches, such as the NCSM.  The aim of this section
is, therefore, to present and partly justify the computation of
effective two-body Hamiltonians acting within a reduced Hilbert space.

The starting point
is a translationally invariant two- or three-body realistic interaction.  The first method 
discussed is based on  a diagonalization of the two- or three-body Schr\"odinger
equation in a large harmonic oscillator (HO) basis. In the two-body
case, this basis typically consists of several hundreds of HO
shells. Via a similarity transformation, one obtains  an
effective two-body interaction, acting within a reduced model space used in NCSM.  Alternatively, one can diagonalize the Schr\"odinger
equation in a full momentum space and then carry out  a similarity transformation to a smaller space defined by some momentum cutoff  $\Lambda$. This leads to the so-called low-momentum
  approach, or  $V_{\mathrm{low-k}}$, discussed below. We discuss
  also the so-called similarity renormalization group transformations.
\subsubsection{Similarity transformed effective interactions in
an oscillator basis
\label{sec:LS}}
The translationally invariant Hamiltonian for an $A$-nucleon system,
used in NCSM studies, is %
 \begin{equation}
\label{eq:ham}
 H=\left[\sum_{i=1}^A\frac{{\bf p}_i^2}{2m} -\frac{{\bf
       P}^2}{2mA}\right] +\sum_{i<j}^A V_{NN}^{ij} +\sum_{i<j<k}^A
 V_{NNN}^{ijk},
 \end{equation}
 $V_{NNN}^{ijk}$ is the three-nucleon force, and ${\bf
   P}=\sum_{i=1}^A{\bf p}_i$ is the center-of-mass momentum.  Since we
 want to employ a HO single-particle basis, it is convenient to use
 the relation
 \begin{equation}
 \sum_{i=1}^A\frac{1}{2}m\Omega^2{\bf r}_i^2-
 \frac{m\Omega^2}{2A}\left[A^2{\bf R}^2+\sum_{i<j}({\bf r}_i-{\bf
     r}_j)^2\right]=0,
 \end{equation}
 where ${\bf R}={1}/{A}\sum_{i=1}^{A}{\bf r}_i$ is the center-of-mass
 coordinate and $\Omega$ is the oscillator frequency.  Inserting this
 relation, we can rewrite the above Hamiltonian as
 $H(\Omega)=H_0+H_I-H_{\mathrm{CM}}$, where 
 $H_0$ is the HO Hamiltonian, $H_I$ is the interaction term defined by
\begin{equation}
 H_I=\sum_{i<j}^A \left[ V_{ij}-\frac{m\Omega^2}{2A}({\bf r}_i-{\bf
     r}_j)^2\right] +\sum_{i<j<k}^A V_{NNN}^{ijk},
\end{equation}
and  $H_{\mathrm{CM}}= {\bf
  P}^2/{2mA}+mA\Omega^2{\bf R}^2/{2}$ is the center-of-mass term.

Shell-model calculations are carried out in a model space defined by a
projector $P$.  The complementary space to the model space is defined
by the projector $Q=1-P$.
With the above Hamiltonian, we can then construct an effective
interaction acting within the model space $P$, reproducing exactly
$N_P$ eigenvalues of the full Hamiltonian. This can be accomplished by
a similarity (Lee-Suzuki) transformation \cite{ls1980,suzuki1982,suzuki1994,suzuki1995}.  
However, no unique unitary transformation exists: one can
construct infinitely many different unitary transformations which
decouple the model and  complementary ($Q=1-P$) subspaces, as discussed  in
Ref.~\cite{kvaal2008}.  Calculation of the exact $A$-body effective
interaction is, however, as difficult as finding the full space
solution.  Using two-body interactions, the effective interaction is
often approximated by a two-body effective interaction determined from
a two-nucleon subsystem.

The above-mentioned transformation can be performed also in a
three-nucleon space. This will generate effective 3NFs even
if the starting Hamiltonian includes only two-body terms. These
effective interaction calculations are performed in a Jacobi coordinate
HO basis \cite{navratil2000,navratil2009}. An effective three-body
interaction can be computed also for starting Hamiltonians with pure NN
interaction terms. This procedure has been shown to speed up
convergence~\cite{navratil2000}, although it should be noted that all
cluster-approximated effective interactions ($a < A$) will in this
approach reproduce the free interaction results in the infinite model
space limit.
\subsubsection{Similarity transformed effective interactions in
momentum space%
\label{sec:Vlowk}}
Alternatively, instead of carrying out the similarity transformation
in an oscillator basis, one can perform the transformation in momentum
space~\cite{bogner2010}.  This
approach consists of  two steps: (i) a diagonalization of the two-body Schr\"odinger equation in the full momentum space,  and (ii) a similarity transformation \cite{suzuki1994,suzuki1995} to relative momenta $k\in [0,\Lambda]$ fm$^{-1}$, with $\Lambda$ defining the relative momenta model space. Typical values of $\Lambda$ are in the range of $\sim 2$ fm$^{-1}$. This evolution to lower $k$-values  can be performed using renormalization group (RG) methods~\cite{epelbaum1999,bogner2003}. We note, however, that the RG approach differs from the Lee-Suzuki transformation as the $Q$-space block of the effective Hamiltonian is now set to zero. The evolution to lower cutoffs shifts contributions from the sum over intermediate states to the interactions, just as RG equations in quantum field theory shift strength from loop integrals to coupling constants.  The evolved low-momentum effective NN interactions have been dubbed $V_{\mathrm{low-k}}$ \cite{bogner2010}. These interactions are significantly softer; hence, more perturbative, as seen, e.g., in nuclear matter calculations~\cite{bogner2005}. By construction,  $V_{\mathrm{low-k}}$ preserves two-nucleon observables for relative momenta up to the cutoff. It should be noted, however, that no tractable approach to evolve many-body forces in this approach has been proposed so far.

\subsubsection{Similarity renormalization group transformed
  interactions%
\label{sec:SRG}}
In recent years, a new approach to perform the similarity
transformation has been developed and used, in particular, with the
chiral interactions. The similarity renormalization group (SRG)
approach~\cite{glazek1993,wegner1994} builds on the general principle
of RG theory, namely that the relevant details of high-energy physics
for calculating low- energy observables can be captured in
scale-dependent coefficients of operators in a low-energy
Hamiltonian. Using this principle as a guide, the 
transformed  Hamiltonian $H_s = U^\dagger_s H U_s$ can be expressed through
a differential (flow) equation
\begin{equation}
\frac{dH_s}{ds} = \left[ \eta_s, H_s \right],
\end{equation}
in which $\eta_s$ is the generator of the flow, an antihermitian
operator that is related to the unitary transformation at the
resolution scale $s$ through
\begin{equation}
\eta_s \equiv \frac{dU_s}{ds}U_s^\dagger \equiv -\eta_s^\dagger.
\end{equation}
The evolved Hamiltonian is $H_s \equiv T_\mathrm{rel} + V_s$, where
the $s$-dependence is contained fully in the effective
potential $V_s$.  The crucial step of the method is the choice of the most
appropriate generator $\eta$. The
antihermitian property is automatically fulfilled by 
\begin{equation}
\eta_s = \left[ G_s, H_s \right],
\end{equation}
where $G_s$ is a momentum-diagonal operator. With this choice,  the transformed operator flows towards a band-dagonal form in momentum space, thus
achieving the desired decoupling property. Applications to nuclear
forces have used $G_s = T_\mathrm{rel}$~\cite{bogner2007,bogner2010}.  

We should note that the generator of the flow is a many-body operator,
which implies that it  induces many-body terms in the effective
operator, see Sec.~\ref{subsec:lightnuclei}. In principle, the evolution of
many-body forces is straightforward, which is an advantage of the SRG
method. However, the associated technical difficulties are significant.
While the two-body evolution is usually performed in momentum
space~\cite{bogner2007}, the three-body evolution was first
implemented in a three-body HO basis~\cite{jurgenson2009,roth2011}. A
first implementation of the SRG flow equation in three-body momentum
space was first performed for a one-dimensional
model~\cite{akerlund2011} and, more recently, for realistic chiral
interactions~\cite{hebeler2012}.

Finally, we want to highlight an important difference between
the Lee-Suzuki transformation and the
scale-dependent effective interactions discussed in
Secs.~\ref{sec:Vlowk} and \ref{sec:SRG}. While the Lee-Suzuki
transformation also induces many-body terms it will, by construction,
reproduce the bare-interaction results already in a two-body cluster
approximation (i.e., without induced many-body forces) -- as the
many-body space is extended towards the full Hilbert space. This
property is not shared by the SRG and $V_{\mathrm{low-k}}$ effective
interactions. By truncating the flow at two- or three-body level, the
unitarity of the transformation is violated, and the bare result will
not be reproduced, even when the many-body model space becomes the full
Hilbert space.

\subsection{Configuration interaction and coupled cluster theory} 

Present shell-model codes can reach dimensions of $d\sim
10^{10}$ basis
states~\cite{caurier2005,dean2004b,horoi2007,maris2011}, and Monte
Carlo-based shell-model codes can attack problems with 
$d\sim 10^{15-16}$
\cite{kdl1997,mizusaki1996,honma1996,utsuno1999,booth2010}.  Although
these numbers are  impressive, the dimension limitation  has
important implications for calculations of nuclei that involve weakly
bound states and/or resonances, as such states require still larger
basis sets.  Extensions of the NCSM to weakly bound systems will be
discussed below.

Large-scale diagonalization is widely used in many areas of physics,
from quantum chemistry~\cite{helgaker} to nuclear
physics~\cite{caurier2005}. The method is based on a projection of the
model Hamiltonian onto a finite-dimensional subspace of the many-body
Hilbert space; hence, the method is an instance of the
Rayleigh-Ritz method. In the standard-shell model approach, one takes
the stance that the many-body Hamiltonian is composed of two parts:
$\hat{H}_0$ and $\hat{H}_I$, treating the latter as a perturbation of
the former.  The eigenfunctions of the mean-field Hamiltonian,
$\hat{H}_0$, are assumed to comprise a single-particle basis for the
Hilbert space. This leads to a matrix diagonalization problem, hence
the name of the method. As $\hat{H}_I$ is the residual
interaction that ``perturbs''  simple configurations of $\hat{H}_0$, the method is commonly referred to as  the
configuration-interaction method.

The NCSM method differs from the standard shell model in several ways:
(i) there is no core, which implies that all nucleons are included in
the many-body basis; (ii) no mean field is introduced and all
interactions are included explicitly; (iii) the 
center-of-mass motion is separated exactly; and   (iv) NCSM Hamiltonian contains realistic two-
and three-body nuclear interactions and the
Hamiltonian matrix is usually built from these interactions using the
similarity transformations.

Coupled cluster theory employs a different approach to systematically
build the many-body wave functions using a  large number of
single-particle states. In actual calculations one typically limits 
the number of active  many-body states to at most three-particle-three-hole
excitations, whereas the single-particle basis can easily be extended
to some 20 major oscillator shells.  This has important consequences
for studies of weakly bound systems. The coupled-cluster method has a
rich history in both nuclear
physics~\cite{coester1958,kuemmel1978,heisenberg1999,dean2004} and
quantum chemistry~\cite{helgaker,bartlett2007}.

In this contribution we highlight recent achievements and
important lessons learned from {\em ab initio} calculations performed with
the NCSM and coupled-cluster methods. In the following, we 
outline the essential features of these two approaches.

\subsubsection{No-core shell-model theory}\label{sec:ncsm}

The starting point for NCSM calculations is the translationally
invariant Hamiltonian $H$ of Eq.~\eqref{eq:ham}. This Hamiltonian is
however modified by adding a HO center-of-mass Hamiltonian
$H_\mathrm{CM} = T_\mathrm{CM} + U_\mathrm{CM}$, where $U_\mathrm{CM}
= A m \Omega^2 \mathbf{R}^2 / 2$. In practice, the Hamiltonian $H +
H_\mathrm{CM}$ is used as input to the computation of a similarity
transformed effective interaction. At the stage of constructing the
many-body Hamiltonian matrix, the HO center-of-mass term is subtracted
and a Lawson projection term $\beta \left( H_\mathrm{CM} - \frac{3}{2}
\hbar\Omega \right)$ is added to shift spurious CM excitations up in
the energy spectrum.

The truncation of the many-body model space is usually performed in
terms of the total energy. That is, the sum of HO excitations that is
contained in a Slater determinant basis state is restricted according
to $\sum_i^A (2n_i + l_i) \le N_\mathrm{tot,max}$. More often,
however, the basis truncation is specified in terms of
$N_\mathrm{max}$ that measures the total number of HO excitations
above the unperturbed ground state. For $A=3,4$,  these two measures are
the same, but for $p$-shell systems they differ, e.g., for
\nuc{6}{Li}, $N_\mathrm{max} = N_\mathrm{tot,max} - 2$, and for
\nuc{11}{Li}, $N_\mathrm{max} = N_\mathrm{tot,max} - 7$, etc. As an
example, the \nuc{6}{Li} results that are presented in
Sec.~\ref{subsec:lightnuclei} were obtained in model spaces up to
$N_\mathrm{max} = 16$. With this $N_\mathrm{max}$-truncation, the
highest single-particle state that can be reached has the energy
$(2n+l) \hbar\Omega = 17 \hbar\Omega$, which gives 171 possible $nlj$
orbitals and 2280 individual single-particle states ($nljm$). The
dimension of the many-body space is $d = 0.79 \cdot 10^9$. This
Hamiltonian matrix is very sparse and can be diagonalized using
iterative Lanczos 
algorithms for real, symmetric matrices \cite{golub1996}.  The ground-state is usually reached
with $~10-20$ Lanczos iterations. The method provides excitation
spectra as well as wave functions, from which various observables can
be computed.

Since NCSM employs HO basis and the total
energy truncation (including \emph{all} allowed configurations),
any eigenstate of the translationally invariant Hamiltonian 
factorizes into a product of a wave function depending on intrinsic
coordinates, and a wave function depending on the center-of-mass
coordinate. The use of any other single-particle basis, or any other
truncation scheme,  results in a mixing of intrinsic and
center-of-mass motions.

Several important ingredients are not considered here, such as the
construction of antisymmetric, few-body states in the
Jacobi-coordinate basis needed for the computation of interaction
matrix elements. For the many-body systems considered here, the
antisymmetric many-body states are constructed in the uncoupled $M$-scheme.  For a much more complete description of the NCSM method, see Ref.~\cite{navratil2009}.

\subsubsection{Coupled cluster theory \label{sec:cc}} 
The single-reference coupled cluster theory is based on the
exponential ansatz for the ground-state wave function of the
$A$-nucleon system,
\begin{equation}
|\Psi_{0}\rangle = e^{T} |\Phi_0\rangle,
\end{equation}
where $|\Phi_0\rangle$ is an uncorrelated, closed shell reference state,
and $T$ is a linear expansion in particle-hole excitation amplitudes.

The approximation that is made in the coupled-cluster approach is
the truncation of the correlation operator $T$ at a given (low-order)
particle-hole excitation level.  Note that in contrast to the full
configuration interaction method, where the expansion in particle-hole
excitation amplitudes is linear, the expansion is non-linear in the
coupled-cluster approach due to the exponentiation of $T$. The most
commonly used approximation in the coupled-cluster approach is the
truncation of the operator $T$ at the singles- and doubles excitation
level (CCSD). Higher accuracy can be obtained by including triples
excitations (CCSDT) ~\cite{shavittbartlett2009,bartlett2007}. In
terms of computational cost, the CCSD method
scales as $n_o^2n_u^4$, while the full CCSDT scales as $n_o^3n_u^5$, where
 $n_0$ represents the number of occupied orbitals and $n_u$ the
number of unoccupied single-particle states. Since coupled cluster theory
with inclusion of full triples CCSDT is usually considered to be too
computationally expensive,  several approximations to the
solution of the CCSDT equations have been developed.  A sophisticated
way of approximating CCSDT is known as the $\Lambda$-CCSD(T) approach~\cite{kucharski1998,taube2008}, in which the left-eigenvector solution of the CCSD
similarity-transformed Hamiltonian is utilized in the calculation of a
non-iterative triples correction which scales as $n_o^3n_u^4$.
Excited states and neighbors of closed shell nuclei can be computed
within the equation-of-motion methods~\cite{bartlett2007}. Recently,
the coupled-cluster method in an angular momentum coupled scheme was
developed in Ref.~\cite{hagen2008,hagen2010a}. This
allows to address medium mass and neutron rich nuclei starting from
``bare'' chiral interactions.  Results presented in this contribution
will focus on the CCSD and the $\Lambda$-CCSD(T) approach, with
various equation-of-motion methods for open-shell nuclei and excited
states~\cite{hagen2010a,jansen2011}.

\subsection{Theoretical treatment of open systems}
\label{sec:weakbound}
\begin{figure}[hbt]
\begin{center}
\includegraphics[width=0.6\textwidth]{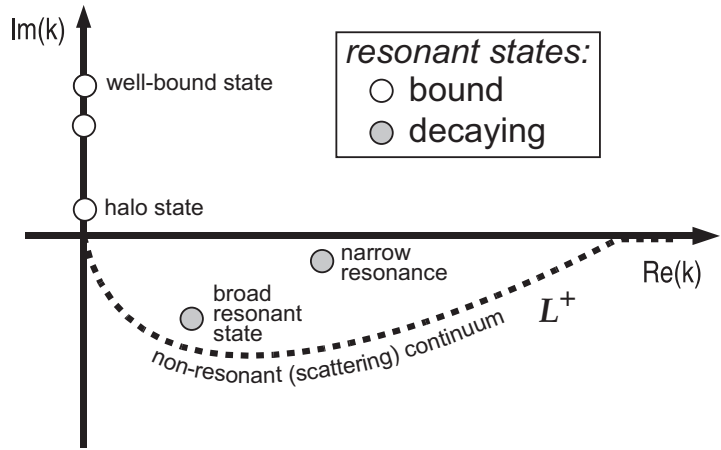}
\caption{Illustration of the Berggren completeness relation
  (\protect\ref{Ber_com}) in the complex $k$-plane. The bound states
  are located on the positive imaginary axis. The weakly bound halo
  states lie close to the origin. The positive-energy resonant states
  are located in the fourth quadrant. Those with a small imaginary
  part can be interpreted as resonances.  The complex-$k$ contour
  $L^{+}$ represents the non-resonant scattering continuum.}
\label{k_plane}
\end{center}
\end{figure}

As discussed earlier, the description of loosely bound and unbound nuclear
states represents a challenge for nuclear structure models
\cite{OpenQS}. Several approaches to this problem are based on the use
of the so-called  Berggren basis \cite{berggren1968,Beggren1993,lind1993}.
Modern applications of the continuum shell model include the
(real-energy) Shell Model Embedded in the Continuum
~\cite{bennaceur1999,okolowicz2003,volya2005a,rotureau2005} based on
the Feshbach projection formalism \cite{feshbach1964}, and the
(complex energy) GSM \cite{michel2002,michel2003,idbetan2003,GSMrev}
employing the Berggren basis \cite{berggren1968,Beggren1993,lind1993}.
The Berggren completeness relation can be written as:
\begin{eqnarray}
\sum_{n=b,d}|\tilde{u}_n\rangle \langle u_n|+\int_{L^{+}}dk
|\tilde{u}_{k}\rangle \langle u_{k}|=1,
\label{Ber_com}
\end{eqnarray}
where $b$ are bound states, $d$ - decaying resonant states, and the
integral along a contour $L^{+}$ in the complex-$k$ plane represents
the contribution from the non-resonant scattering continuum, see
Fig. \ref{k_plane}. In general, different contours can be used for
different $(\ell,j)$ partial waves.

The recent GSM applications include the analysis of threshold effects
in multichannel coupling and spectroscopic factors
\cite{cusps,sfacts}, description of isospin mixing in weakly bound
nuclei \cite{Isospinmix}, explanation of behavior of charge radii in
helium halo nuclei \cite{papadimitriou2011}, and studies of asymptotic
normalization coefficients in mirror nuclei \cite{ANCGSM}. Another
exciting development utilizing the Berggren basis is the extension of
the coupled-cluster method to open systems; it represents the first
complex-energy nuclear {\em ab initio} framework
~\cite{hagen2007d,hagen2012a, hagen2012b} capable of describing
many-body bound and unbound states.  In actual coupled-cluster
calculations, one performs first a Hartree-Fock calculation and
transforms the Hamiltonian to the Hartree-Fock basis. As a
consequence, the coefficients $t_i^a$ of the singles amplitude, see
Sec.~\ref{sec:cc}, acquire small values in the solution of the coupled-cluster 
equation. One also finds, when using a HO basis, that this
approach reduces the $\hbar\Omega$-dependence of the computed
energies. In practical computations one aims at increasing the number
$N+1$ of employed oscillator shells until the results become virtually
independent on the model-space parameters.  In Ref.~\cite{hagen2006}
it was demonstrated that one could start with a standard HO basis and
corresponding interactions. We refer the reader to the latter
reference for further details.

In order to use the Berggren basis in large-scale configuration-mixing
calculations, the integral over the non-resonant continuum is
discretized by using a suitable quadrature rule. Using this discrete
Berggren basis, the GSM basis is obtained in the usual way by
constructing many-body Slater determinants.  The dimension of the
Hamiltonian matrix grows rapidly with the number of discretized
continnum states and the number of nucleons; hence, advanced numerical
methods that can handle large non-Hermitian matrices must be used. In
the context of the GSM, it has been shown that the Density Matrix
Renormalization Group (DMRG) is an efficient way to compute the
low-lying spectrum of the Hamiltonian at a low computational cost.

\subsubsection{Density Matrix Renormalization Group for the GSM%
\label{sec:dmrg}}
The DMRG method was first introduced to overcome the limitations of
the Wilson-type renormalization group to describe strongly correlated
systems with short-range interactions \cite{white1992,white1993}. More
recently, the DMRG has been reformulated and applied to finite Fermi
systems \cite{dukelsky2002}, nuclear shell model
\cite{pittel2006,papenbrock2005,thakur2008}, and open systems
\cite{rotureau2006}. While most of the DMRG studies have been focused
on properties in strongly correlated closed quantum systems
characterized by Hermitian density matrices, systems involving
non-Hermitian and non-symmetric density matrices can also be treated
\cite{carlon1999,rotureau2006,rotureau2009}.

\begin{figure}[hbt]
\begin{center}
\includegraphics[width=0.3\textwidth]{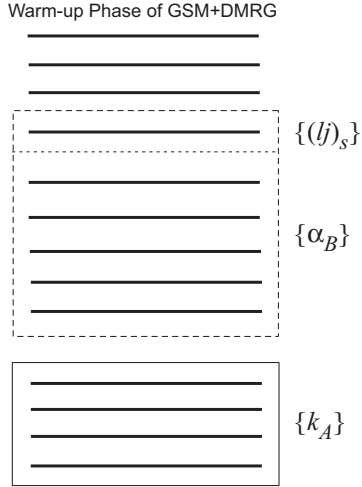}
\caption{Schematic illustration of the GSM+DMRG procedure during the
  $s^{th}$ step of the warm-up phase.  States $\{k_{A} \}$ from $A$,
  previously optimized states $\alpha_{B} $, and states $\{(lj)_s \}$
  constructed by occupying the $s^{th}$ shell with $n$ particles are
  coupled to generate the new set of states $\{k_A \otimes i_B\}^J=
  \{k_{A} \otimes \{\alpha_B \otimes (lj)_s^n \} \}^J $.}
\label{fig:warmup}
\end{center}
\end{figure}

Let us consider the application of the $J$-scheme DMRG in the context
of the GSM (GSM+DMRG). The objective is to calculate an eigenstate
$|J^{\pi}\rangle$ of the GSM Hamiltonian $\hat{H}$ with angular
momentum $J$ and parity $\pi$.  As $|J^{\pi}\rangle$ is a many-body
pole of the scattering matrix of $\hat{H}$, the contribution from
non-resonant scattering shells along the continuum contour $L^{+}$ to
the many-body wave function is usually smaller than the contribution
from the resonant orbits \cite{GSMrev}. Based on this observation, the
following separation is usually performed \cite{rotureau2006}: the
many-body states constructed from the single-particle poles form a
subspace $A$ (the so-called `reference subspace'), and the remaining
states containing contributions from non-resonant shells form a
complement subspace $B$ (see Fig. \ref{fig:warmup}).

One begins by constructing states $|k\rangle_A$ forming the reference
subspace $A$.  All possible matrix elements of suboperators of the GSM
Hamiltonian $\hat{H}$ acting in $A$, expressed in the second
quantization form, are then calculated and stored and the GSM
Hamiltonian is diagonalized in the reference space to provide the
zeroth-order approximation $|\Psi_J\rangle^{(0)}$ to
$|J^{\pi}\rangle$. This vector, called `reference state', plays an
important role in the GSM+DMRG truncation algorithm.
The scattering shells $(lj)$, belonging to the discretized contour
$L^{+}$, are then gradually added to the reference subspace to create
the subspace $B$. This first stage of the GSM+DMRG procedure is
referred to as the 'warm-up phase.  For each new shell that is added,
all possible many-body states denoted as $|i\rangle_B$ are constructed
and matrix elements of suboperators of the GSM Hamiltonian acting on
$|i\rangle_B$ are computed.
By coupling states in $A$ with the states $|i_B\rangle$, one
constructs the set of states of a given $J^{\pi}$. This ensemble
serves as a basis in which the GSM Hamiltonian is diagonalized.  The
target state $|\Psi_J\rangle$ is selected among the eigenstates of
$\hat{H}$ as the one having the largest overlap with the reference
vector $|\Psi_J\rangle^{(0)}$. Then, the desired truncation is
performed in $B$ by introducing the reduced density matrix,
constructed by summing over the reference subspace
$A$~\cite{mcculloch2002}.  In standard DMRG applications for Hermitian
problems, where the eigenvalues of the density matrix are real
non-negative, only the eigenvectors corresponding to the largest
eigenvalues are kept during the DMRG process.  Within the metric
defining the Berggren ensemble, the GSM density matrix is
complex-symmetric and its eigenvalues are, in general, complex.  As a
consequence, the truncation is done by keeping the eigenstates
$\alpha_{B} $ (the `optimized' states) with the largest nonzero moduli
of eigenvalue~\cite{rotureau2009}.

The warm-up phase is followed by the so-called sweeping phase, in
which, starting from the last scattering shell $(lj)_\mathrm{last}$,
the procedure continues in the reverse direction (the `sweep-down'
phase) until the first scattering shell is reached. The procedure is
then reversed and a sweep in the upward direction (the 'sweep up'
phase) begins. The sweeping sequences continue until convergence for
target eigenvalue is achieved.

Applications of GSM+DMRG have been reported in Refs.
\cite{rotureau2006,rotureau2009,papadimitriou2011}. Those examples
demonstrate that weakly-bound and unbound nuclear systems, which --
because of a prohibitively large size of Fock space, cannot be treated
by means of direct diagonalization techniques -- can be treated very
efficiently using DMRG. The DMRG also opens up the avenue to perform
NCSM calculations for open systems using the Berggren
basis~\cite{papadimitriou-unpub}.

\subsubsection{The Resonating Group Method%
\label{sec:rgm}}
An alternative formulation for describing open quantum systems,
characterized by a limited number of open channels, is the resonating
group method (RGM)~\cite{wildermuth1977}. Here, the many-body wave
function is decomposed into contributions from various channels that
are distinguished by their different arrangement of the nucleons into
clusters.  In principle, this corresponds to the expansion of the
bound state, or the interior region of a scattering state, into an
over complete set of basis functions. The basis functions consist of
two parts: the cluster wave functions and the wave function
representing the relative motion of the clusters. In the case of two
clusters, for instance, the full wave function of the system can be
written as
\begin{equation}
\Psi^{(A)} = \sum_\nu \mathcal{A}_\nu \Big\{ \Phi_{1\nu} \Phi_{2\nu}
\varphi_\nu (\vec{r}_\nu) \Big\},
\end{equation}
where $\nu$ labels cluster channels.  This expansion is complicated
due to the presence of the antisymmetrizer $\mathcal{A}_\nu$, which
accounts for the exchange of nucleons between the clusters. The
intrinsic wave functions, $\Phi_{1,2}$, are internally antisymmetric
and would, in the NCSM+RGM
approach~\cite{quaglioni2008,quaglioni2009}, be eigenstates of the
NCSM Hamiltonian for that particular cluster.

At this point, a basis of binary-cluster channel states, $|\Phi^{J^\pi
  T}_{\nu r}\rangle$, is introduced that includes the spin-coupled
product of internal (antisymmetric) wave functions of the two clusters
at relative distance $r$.  This basis can be used to expand the
many-body wave function:
\begin{equation}
|\Psi^{J^\pi T}\rangle = \sum_{\nu} \int dr \,r^2\frac{g^{J^\pi
    T}_\nu(r)}{r}\,\hat{\mathcal A}_{\nu}\,|\Phi^{J^\pi T}_{\nu
  r}\rangle\,.
\label{eq:trial}
\end{equation}
By diagonalizing the NCSM Hamiltonian in this basis, one obtains the
RGM equations:
\begin{equation}
\sum_{\nu}\int dr \,r^2\left[{\mathcal H}^{J^\pi
    T}_{\nu^\prime\nu}(r^\prime, r)-E\,{\mathcal N}^{J^\pi
    T}_{\nu^\prime\nu}(r^\prime,r)\right] \frac{g^{J^\pi T}_\nu(r)}{r}
= 0,
\label{RGMeq} 
\end{equation}
where ${\mathcal N}^{J^\pi T}_{\nu^\prime\nu}(r^\prime, r)$ and
${\mathcal H}^{J^\pi T}_{\nu^\prime\nu}(r^\prime, r)$ are the norm and
Hamiltonian kernels, respectively.  These non-local quantities contain
all the nuclear structure information. However, we note that the basis
states are asymptotically orthogonal so that all important physical
quantities (such as the scattering matrix) can be defined with the
asymptotic solution.  The non-orthogonality mainly appears at short
distances and is primarily due to antisymmetrization effects. The RGM
equations can be solved by means of the standard symmetric
orthogonalization method~\cite{quaglioni2009}.

The NCSM+RGM approach has been initially used for single-nucleon
projectiles, both for scattering and bound-state
cases~\cite{quaglioni2008,quaglioni2009,navratil2010}. The method has
recently been extended to study projectile-target binary-cluster
states, where the projectile is a deuteron~\cite{navratil2011}. This
new development makes it possible to compute, e.g., $d +$\nuc{4}{He}
scattering, clusterized \nuc{6}{Li} states, as well as deuteron fusion
reactions on \nuc{3}{H} and \nuc{3}{He}~\cite{navratil2012}.

\section{Selected results}\label{sec:results}

In this section, we focus on selected results for several chains of
neutron-rich isotopes, both light and heavy. We base our
theoretical analyses on large-scale NCSM calculations, coupled-cluster
calculations, and nuclear DFT. An important
question is the role played by three
or more complicated many-body terms in the low-energy effective
nuclear Hamiltonian.  The study of 3NFs in medium mass and neutron
rich nuclei is an ongoing research topic in nuclear many-body
physics.  Several calculations indicate  that 3NFs are needed to understand various properties of nuclei, see for example
Refs.~\cite{hagen2012a,hagen2012b,otsuka2010,navratil2009,wiringa2002,epelbaum2011,maris2011,roth2011,roth2012}.
As one moves to more neutron-rich isotopes, correlations play an
increasingly more important role. The mean field contribution is often
reduced relatively to the effects due to   two-, three- and many-body correlations.  Furthermore,
close to the drip lines, the number of degrees of freedom increases
dramatically, with resonant and non-resonant continuum channels
becoming available. All of this requires a good understanding of correlations and interactions driving the  observed properties.

A three-body force is expected to play a  role in
the evolution of single-particle energies as more and more 
particles are added. For instance, the effect of the monopole term has been analyzed
intensively in terms of phenomenological interactions over the last
two decades~\cite{taka2001a,taka2005,honma2002,caurier2005}. However,
there is also clear evidence from several calculations that a traditional
shell-model picture with a strong mean field that defines a 
single-particle basis, may break down, see for example
Refs.~\cite{NazPPNP,hagen2012b} and discussion below.

As outlined in this contribution, we are now in a position where 3NFs
from EFT can be included routinely into various {\em ab initio}
methods. This means that we can start to explore, which components
of the nuclear forces are at play when we move towards the drip lines.
Furthermore, with long chains of isotopes to be studied
theoretically and experimentally, one can
eventually attempt to extract in-medium information that will allow us
to constrain 3NFs for heavier nuclei.

In the following, selected results for light and medium-heavy are discussed. We shall focus on results for binding energies and a
few excited states for selected chains of isotopes In several cases we
will study the impact of  3NFs and couplings to the continuum.

\subsection{Results for light nuclei%
\label{subsec:lightnuclei}}
In this section,  we  mainly focus on {\em ab
  initio} description of light nuclei near the drip lines. We shall
 restrict discussion to approaches that try to solve the many-body
Schr{\"o}dinger equation without uncontrolled approximations, using a
well-defined microscopic Hamiltonian as a single input. While there
are several realistic nuclear interactions that reproduce NN
scattering phase shifts and few-body data with  high precision, we focus mainly
on results obtained with chiral interactions.

Bound state {\em ab initio} calculations of light nuclei were pioneered by
the Variational Monte Carlo (VMC) and Green's Function Monte Carlo
(GFMC) methods~\cite{pieper2001a}. They have studied $A=4-12$
systems using various combinations of Argonne NN interactions and UIX
or Illinois (IL) 3NFs, and have demonstrated an impressive agreement
with experimental energies, not only for bound states but also for
narrow resonances~\cite{leidemann2012}.

{\em Ab initio}  methods that can currently model systems with $A>4$, 
include the
Effective Interaction for Hyperspherical Harmonics
(EIHH)~\cite{barnea2000,barnea2001}, lattice calculations
\cite{epelbaum2011,epelbaum2010}, the coupled-cluster
method~\cite{dean2004,hagen2007c,hagen2010a}, and the No-Core Shell
Model~\cite{navratil2000b,navratil2000c,navratil2009}. The EIHH is
currently limited to $A \leq 6$ due to the difficulty of
antisymmetrization. Calculations on the lattice for light bound
systems~\cite{epelbaum2010} and the Hoyle state~\cite{epelbaum2011}
have been recently reported.

As a benchmark example, and an illustration of the use of modern
chiral interactions including consistent 3NFs in NCSM, we present
\begin{table*}[hbt]
\caption{Properties of \nuc{3}{H} and \nuc{4}{He}. Benchmarking of
  calculations with chiral NN+3NF interactions using
  NCSM~\cite{navratil2007a} and Hyperspherical
  Harmonics~\cite{kievsky2008} methods. (From
  Refs.~\cite{navratil2009,forssen2011}.)}
\label{tab:3h4he}
\centering
\begin{tabular}{lllllll}
\hline\noalign{\smallskip} & & \multicolumn{2}{c}{NN (N$^3$LO)} &
\multicolumn{2}{c}{+3NF(N$^2$LO)} & Expt. \\ & & NCSM & HH & NCSM & HH
\\ \noalign{\smallskip}\hline\noalign{\smallskip} \nuc{3}{H} &
$E_\mathrm{gs}$ [MeV] & 7.852(5) & 7.854 & 8.473(5) & 8.474 & 8.482
\\ & $\langle r_p^2 \rangle^{1/2}$ [fm] & 1.650(5) & 1.655 & 1.608(5)
& 1.611 & 1.60 \\ \noalign{\smallskip}\hline\noalign{\smallskip}
\nuc{4}{He} & $E_\mathrm{gs}$ [MeV] & 25.39(1) & 25.38 & 28.34(2) &
28.36 & 28.296 \\ & $\langle r_p^2 \rangle^{1/2}$ [fm] & 1.515(2) &
1.518 & 1.475(2) & 1.476 & 1.467(13) \\ \noalign{\smallskip}\hline
\end{tabular}
\end{table*}
in Table~\ref{tab:3h4he} results  for $A = 3, 4$ obtained with and without inclusion of 3NFs. The
chiral 3NF is from order N$^2$LO and contains two low-energy constants
(LEC) that need to be determined from $A>2$ data. In
Ref.~\cite{navratil2007a}, these constants were obtained from $A=3$
binding energies, but also by investigating sensitivities of
properties of $A \ge 4$ nuclei to the variation of the constrained
LECs. Besides the triton ground state energy, which is by construction
within a few keV of experiment, the NN+3NF results for the \nuc{4}{He}
ground-state energy and point-proton radius are in excellent agreement
with measurement. We also note a very good agreement between NCSM and  the variational HH
method~\cite{kievsky2008}.

Developments based on an importance-truncation (IT) have recently been
proposed in Ref.~\cite{roth2007} to address the factorial growth of
the NCSM model space with $N_\mathrm{max}$ and particle number A. With
a criterion based on perturbation theory, a large fraction of the
many-body basis states can be discarded and calculations with
$N_\mathrm{max} = 12$ for \nuc{12}{C} and \nuc{16}{O} have become
feasible~\cite{roth2011}.

Let us first consider the chain of He isotopes. The charge radii and
masses of He isotopes (up to \nuc{8}{He}) were recently determined
experimentally~\cite{mueller2007,brodeur2012} and compared to results
from {\em ab initio} methods. For \nuc{6}{He} in particular, results for radii
and binding energies are available from calculations with
GFMC~\cite{pieper2008}, NCSM~\cite{caurier2006}, FMD~\cite{neff2005},
and EIHH~\cite{bacca2009,brodeur2012}. Unfortunately, at the present stage, those results can not be
used for benchmarking as different interactions have been used. Actually, the GFMC results are the only published converged
calculations that include 3NFs. Within the range of 
GFMC results, it is possible to reproduce both the charge radius and
 separation energy. However, there is a large uncertainty due to 
different models of 3NFs  and different trial wave functions used.

\begin{figure}[hbt]
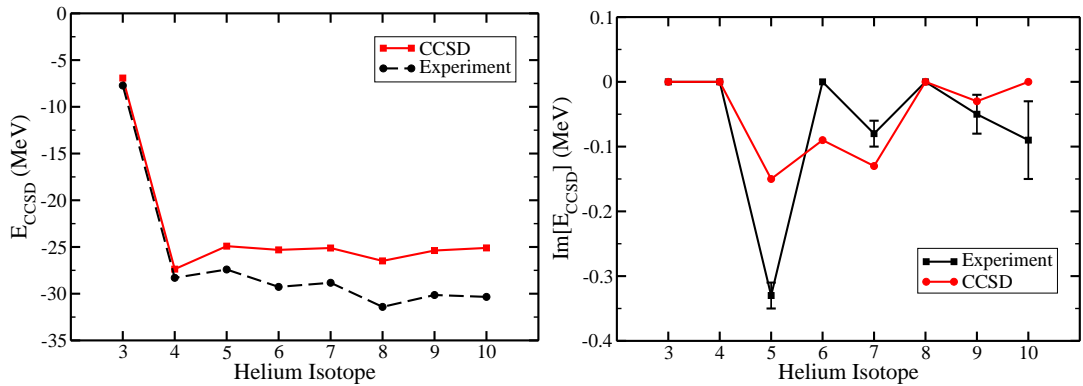

  \centering
  \includegraphics*[width=0.45\columnwidth]{figs/He_Isotopes_CC_ReE.eps}
  \includegraphics*[width=0.45\columnwidth]{figs/He_Isotopes_CC_ImE.eps}
  \caption{Left panel: CCSD ground state energies (red solid line)
    compared to experimental binding energies (black dashed line) for
    \nuc{3-10}{He}.  Right panel: Imaginary parts of the CCSD (red
    solid line) and experimental (black dashed line) ground state
    energies for\nuc{3-10}{He}.}.
  \label{fig:he_egs}
\end{figure}

The first {\em ab initio} coupled-cluster calculations using a Berggren
basis were performed in Ref.~\cite{hagen2007a}. Within the coupled
cluster singles-and-doubles approximation (CCSD), the ground state
binding energies and lifetimes of the \nuc{3-10}{He} isotopes were
calculated employing  $V_{{\mathrm low}-k}$
with cutoff $\Lambda = 1.9 \mathrm{fm}^{-1}$, derived from the N$^3$LO
nucleon-nucleon interaction of Ref.~\cite{entem2003}.
A comparison between the CCSD results and experiment is shown
in Fig.~\ref{fig:he_egs} and demonstrates a fair agreement. An improved description of the helium isotopes
must include the effects of 3NFs and triples correlations (CCSDT).

\begin{figure}[hbt]
\centerline{ \includegraphics[width=0.8\columnwidth,
    clip=true]{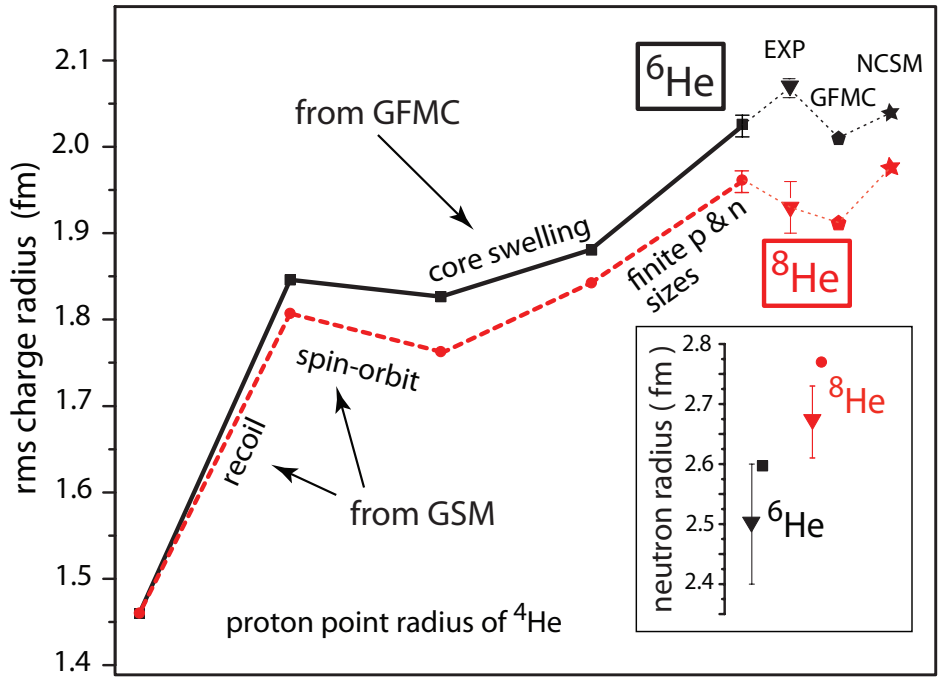} }
  \caption[T]{\label{fig:charge-rad-GSM} Different contributions to
    the charge radius of $^{6}$He (solid line, squares) and $^{8}$He
    (dashed line, dots) calculated in GSM. The core swelling
    contribution is taken from GFMC calculations of
    Ref.~\cite{pieper2001a}. Experimental charge radii come from
    \cite{brodeur2012} (triangles).  The NCSM~\cite{caurier2006}
    (stars) and GFMC~\cite{pieper2008} (pentagons) results are marked
    for comparison. The inset shows GSM rms neutron radii compared to
    experiment~\cite{alkhazov1997}. (From Ref.~\cite{papadimitriou2011}.)}
\end{figure}

A hallmark feature of halo systems appearing near the nucleon drip
line is the rapid increase of matter radii. Corresponding effects
on the charge radius  are much smaller, but this quantity can be studied very
accurately using laser spectroscopy techniques. Let us consider the neutron-rich
nuclei \nuc{6,8}{He} as an example.
The charge radius of \nuc{6}{He} is 2.059(7) fm and exceeds that of
\nuc{8}{He}, which is 1.959(16)~fm, see
Refs.~\cite{mueller2007,brodeur2012}. Both of them are much larger
than the charge radius of \nuc{4}{He}, 1.681(4)~fm. These two heavy
helium isotopes are neutron halo nuclei and, since the protons are
confined inside the tightly bound $\alpha$-core, the differences of
charge radii carry unique structural information on the nuclear
Hamiltonian and  many-body dynamics. In
Ref.~\cite{papadimitriou2011}, \nuc{6}{He} and \nuc{8}{He} were
described with the GSM as systems of valence neutrons moving  outside the
$\alpha$ core, and interacting via a finite-range Minnesota
potential. The core-neutron potential was described by a Woods-Saxon
potential and the model space was constructed from the Berggren
basis. While such a calculation is obviously not not {\em ab initio}, it allows us
to understand intricate experimental data  in simple terms. Specifically, the GSM work demonstrated that the observed
charge radii depend mainly on three factors: (i) the recoil due
to the motion of valence nucleons around the $\alpha$-core; (ii)  the spin-orbit term; and (iii) the swelling of the $\alpha$-core in the neutron environment
(see Fig.~\ref{fig:charge-rad-GSM}). While (i) and (ii) are robustly predicted by GSM (i.e., they depend weakly on the interaction, provided that the threshold energies are under control), the core swelling effect (iii) must be taken from an {\em ab initio} theory (here, from GFMC calculations~\cite{pieper2008}). One can thus conclude that the GSM approach links the high-quality atomic data to a subtle in-medium effect that provides a stringent test of {\em ab initio} theory. 

The chain of lithium isotopes offers many splendid examples of drip line physics. The
nucleus \nuc{6}{Li} belongs to the valley of beta-stability and
\nuc{7,8,9}{Li} are particle-bound but beta-unstable.  Adding one more
neutron to \nuc{9}{Li} yields an unbound \nuc{10}{Li}, while adding a
pair of neutrons produces the additional pairing energy that makes
\nuc{11}{Li} bound. Such an even-odd staggering effect in binding
energy can be often seen along the neutron drip line.
The conventional  shell model picture of \nuc{11}{Li} puts the
valence neutrons in a $p$-wave state. However, measured angular
correlations in fragmentation reactions~\cite{simon2007} demonstrate
the presence of states with different parities. Furthermore, the structure of the unbound \nuc{10}{Li}
is expected to correspond to a virtual $s$-state with a very large
scattering length. These phenomena: melting and re-organizing of shell
structure, ground states embedded in the continuum, and dilute matter
densities such as halos, appear in most isotopes around the drip
lines. They just illustrate a  point that will be made
 clear throughout this contribution, namely, light nuclei living on the
edge of stability cannot be described within a mean-field picture.

The Li and Be isotopic chains were recently studied in the
NCSM~\cite{forssen2009} to investigate the wealth of exotic properties
that generally pose a challenge for nuclear-structure models: (i) 
appearance of clustering; (ii) halo structure of \nuc{11}{Li}; and 
(iii) low ground-state quadrupole moment of \nuc{6}{Li}.
In this systematic study,  series of
calculations were carried out in  large model
spaces using CD-Bonn 2000~\cite{machleidt2001}, and INOY
(IS-M)~\cite{doleschall2004} interactions.   The degree of convergence was
estimated from the \nm- and \ho-dependence of the results.
Alternatively, one can utilize the fact that NCSM calculations
performed at different HO frequencies should all converge to the same
value in the limit $\nm \to \infty$. This constitutes an example of
multiple converging sequences in the NCSM, discussed extensively in
Ref.~\cite{forssen2008}. In this context, we note that much work is
currently focused on quantifying the basis truncation corrections~\cite{coon2012,furnstahl2012}.

\begin{figure}[hbt]
\centering \includegraphics*[width=0.6\columnwidth]
           {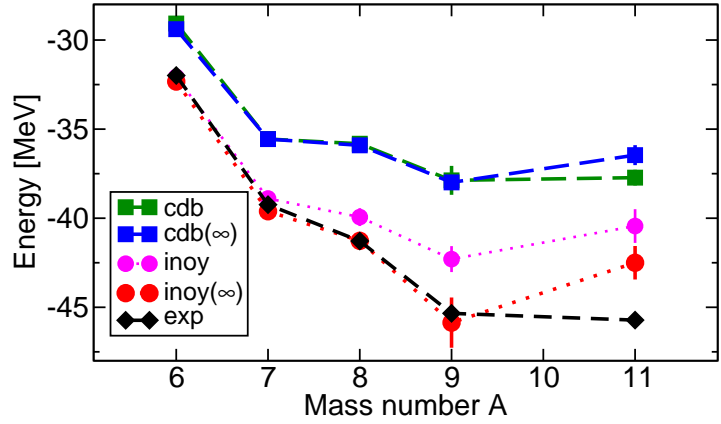}
\caption{Ground-state energies for Li isotopes predicted in NCSM
  compared with experiment. The exponential convergence rate
  was not fully reached for \nuc{11}{Li}. (From
  Refs.~\cite{forssen2009,forssen2011}.)
  \label{fig:li_egs}}
\end{figure}

Figure~\ref{fig:li_egs} shows the ground-state energies of $A=6-11$ Li
isotopes. The isotopic trend is nicely reproduced, but also the known
feature of pure NN interactions giving too little binding for
many-nucleon systems is seen. The INOY interaction is a bit different as the
NN $P$-wave scattering has been modified slightly in order to
reproduce binding energies and the analyzing powers in the $A=3$
systems. Note that the model spaces used in computations of
\nuc{11}{Li} were not large enough to reach the exponential
convergence region.

\begin{figure}[hbt]
\centering
\includegraphics*[width=0.6\columnwidth]{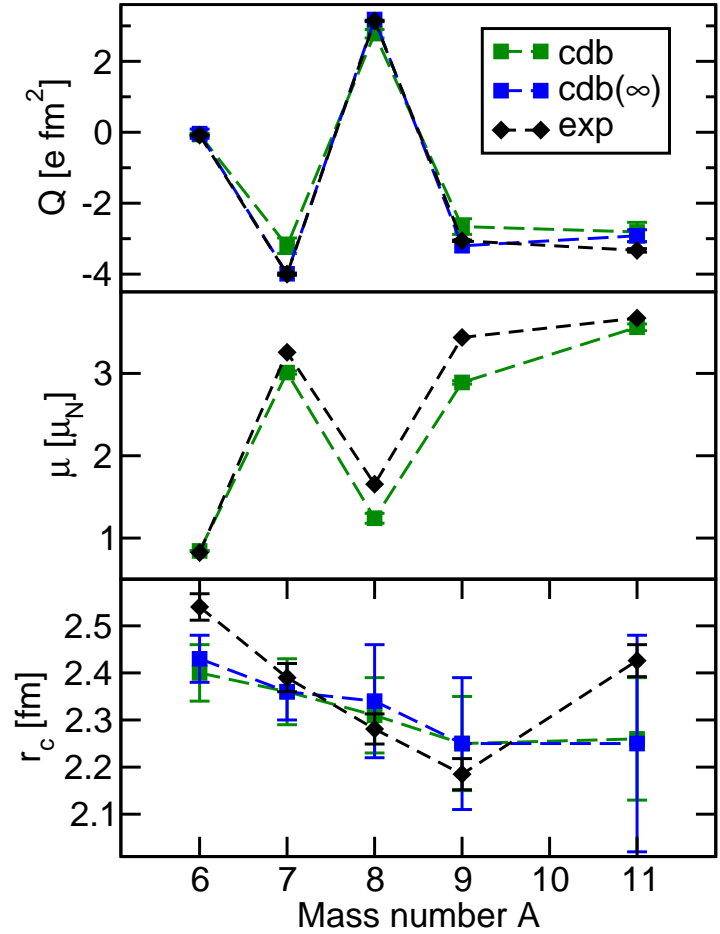}
\caption{Electric quadrupole moments, magnetic dipole
  moments, and charge radii of Li isotopes obtained in NCSM and compared with experiment. (From Refs.~\cite{forssen2009,forssen2011}.)
  \label{fig:liA}}
\end{figure}

In Fig.~\ref{fig:liA} we compare the calculated and experimental
trends for a number of observables for the Li chain of isotopes.
With the important exception of the radius of the \nuc{11}{Li} halo
ground-state, a  good agreement between NCSM results and 
experiment is found. In particular, the overall trends are well
reproduced. In this context, however, it
is important to point out that  calculations of electromagnetic
moments are usually performed in the impulse approximation (i.e., one-body
nucleon currents), and that the inclusion of
meson exchange currents can add about $20\%$ to magnetic dipole
moments~\cite{marcucci2008}. We expect more work in this direction,
including fully consistent two-body currents from chiral Lagrangians.
Another success of NCSM calculations, is the reproduction of the small quadrupole moment of \nuc{6}{Li} that is
known to pose a challenge for  theory. In
particular, the general failure of three-body models for this
observable has been blamed on missing antisymmetrization of the
valence nucleons and the nucleons in the
alpha-core~\cite{unkelbach1991}. In addition, the ratio
$Q\left( ^{11}\mathrm{Li} \right) / Q\left( ^{9}\mathrm{Li} \right)$
is found to be very close to unity, as confirmed by very precise
experimental data~\cite{neugart2008}, and  the trend for the much larger
moments of $A=7-11$ is nicely reproduced. This  has been obtained within
the truncated HO basis space that does not give a very accurate
description of the dilute halo structure of \nuc{11}{Li}. Still, the
observed decrease of the charge radius in $A=6-9$ is reproduced.

A clear example of the disappearance of magic numbers in weakly
bound systems is found in \nuc{11}{Be}. The experimental ground state of
\nuc{11}{Be} is an intruder $1/2^+$ level, while the first $p$-shell
state is $1/2^-$ situated at $E_x = 320$~keV. The neutron separation
energy is only 503~keV, and there are no other bound states.
An investigation of $A=9,11$ isotopes in large-basis {\em ab initio} NCSM
calculations were reported in Ref.~\cite{forssen2005}. Calculations were
performed for both natural-parity and unnatural-parity states in model
spaces up to $\nm = 9$ using four different accurate NN
potentials. The \nuc{11}{B} $\nm = 9$ calculation, with a basis
dimension of $1.1 \cdot 10^9$, was the largest NCSM diagonalization at
that time.
The calculations did not reproduce the anomalous $1/2^+$ ground state, but did predict a dramatic drop in the
positive-parity excitation energies with an increasing model space.
Furthermore, the behavior of the INOY results suggested that a
realistic NNN force might have an influence on the observed parity
inversion.

It is  expected that the halo character of a loosely
bound neutron  should have a major influence on the
characteristics of a nuclear  state.
This fact is illustrated by the extremely strong E1 transition between
the two bound states in \nuc{11}{Be}, that can be explained by a large
overlap of the initial and final state wave functions at large
distances. The E1 strength was underestimated by a factor of 20 in
the NCSM calculations~\cite{forssen2005}, demonstrating that the halo
character of this state is extremely hard to reproduce using a HO
basis.
This observation makes \nuc{11}{Be} an excellent candidate for testing
the NCSM+RGM method, in which the relative motion of the core and the
valence nucleon is treated more accurately. By imposing bound-state
boundary conditions to the set of coupled channel Schr{\"o}dinger
equations of Eq.~\eqref{RGMeq}, the bound states of \nuc{11}{Be} could be
studied in a NCSM+RGM model space spanned by the $n+$\nuc{10}{Be}
channel states with inclusion of the ground state plus three excited
states of \nuc{10}{Be}~\cite{quaglioni2008}. Binding energies of the
various \nuc{10,11}{Be} states are displayed  in Table~\ref{tab:be} for
both  NCSM and  NCSM+RGM. The precise treatment of
the neutron halo  strongly influences the $S$-wave relative
kinetic and potential energies. The rescaling of the relative wave
function in the internal region,
seen in the NCSM+RGM, is the main cause of the dramatic decrease
($\sim 3.5$~MeV) of the energy of the $1/2^+$ state. This effect makes
the $1/2^+$ state bound and even leads to a parity inversion in
 NCSM+RGM.
\begin{table}
\begin{center}
\begin{tabular}{lccclrclr}
&&$^{10}$Be&&\multicolumn{2}{c}{$^{11}$Be($1/2^-$)}&&\multicolumn{2}{c}{$^{11}$Be($1/2^+$)}\\[0.7mm]\cline{3-3}\cline{5-6}\cline{8-9}\\[-4mm]
  &$N_{\rm max}$&$E_{\rm g.s.}$
&&\multicolumn{1}{c}{$E$}&\multicolumn{1}{c}{$E_\mathrm{th}$}&&\multicolumn{1}{c}{$E$}&\multicolumn{1}{c}{$E_\mathrm{th}$}\\[0.5mm]
  \hline NCSM~\cite{forssen2005} & 8/9&
  -57.06&&-56.95&0.11&&-54.26&2.80\\ NCSM~\cite{forssen2005} & 6/7&
  -57.17&&-57.51&-0.34&&-54.39&2.78\\ NCSM+RGM~\cite{quaglioni2008}
  &&&&-57.59&-0.42&&-57.85&-0.68\\ Expt. & &
  -64.98&&-65.16&-0.18&&-65.48&-0.50 \\ \hline
\end{tabular}
\caption{Energies (in MeV) of the ground state of  $^{10}$Be  and 
  the lowest states of $^{11}$Be,
  calculated in NCSM using the CD-Bonn NN potential~\cite{machleidt2001} at
  $\ho=13$~MeV. The NCSM+RGM results were obtained using $n+^{10}$Be
  configurations with $\nm= 6$ g.s., $2^+_1$, $2^+_2$, and $1^+_1$
  states of $^{10}$Be. (From Refs.~\cite{forssen2005,quaglioni2008}.)
\label{tab:be}}
\end{center}
\end{table}

Let us mention also modern nuclear reaction calculations that offer
opportunities to compute new observables that probe other properties
of realistic nuclear Hamiltonians.  Here we present results from $N+\alpha$
scattering using the NCSM+RGM approach. The $A=5$ system is an ideal
testing ground for many-body scattering theory for several reasons: (i)
the $A=5$ system does not have a bound state; (ii) \nuc{4}{He} is
tightly bound so that single-channel scattering is valid up to
$\sim20$~MeV; (iii) there are two low-lying p-wave resonances ($3/2^-$
and $1/2^-$); (iv) non-resonant s-wave scattering ($1/2^+$) for which
large effects of the Pauli exclusion principle is expected; (v) it is a well studied system.  In
particular, there have been recent microscopic studies of $N+\alpha$
scattering  with  GFMC~\cite{nollett2007} and  NCSM+RGM~\cite{quaglioni2008,quaglioni2009}.

\begin{figure}
\centering{\includegraphics[width=0.6\columnwidth]
  {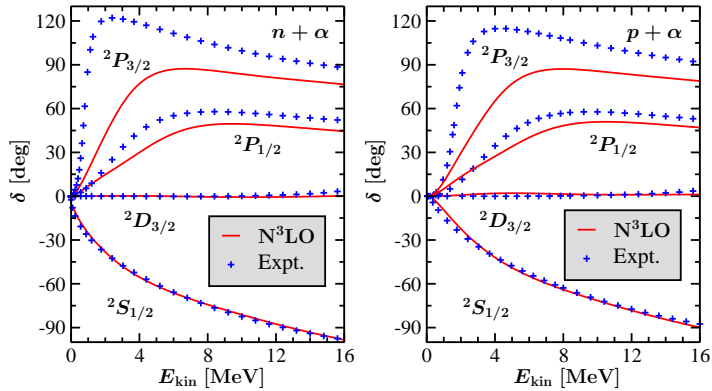}}
\caption{Calculated phase shifts for (left panel) $n$-$\alpha$ and
  (right panel) $p\,$-$\alpha$ scattering, using the N$^3$LO NN
  potential~\cite{entem2003}, compared to an $R$-matrix analysis of
  data ($+$). Theoretical results include the $^4$He g.s., $0^+ 0$,
  $0^- 0$, $1^-0$, $1^-1$, $2^- 0$, and $2^-1$ states. (From
  Refs.~\cite{quaglioni2008,forssen2011}.)
\label{fig:npaexp}}
\end{figure}
A comparison with an accurate $R$-matrix analysis of the
nucleon-$\alpha$ scattering is presented in Fig.~\ref{fig:npaexp}. It
reveals that for both neutron (left panel) and proton (right panel)
projectiles one can describe  well the $^2S_{1/2}$ and,
qualitatively, also the $^2D_{3/2}$ phase shifts, using the N$^3$LO NN
potential. The good agreement of the N$^3$LO $^2S_{1/2}$ phase shifts
with the $R$-matrix analysis can be credited to the repulsive action
of the Pauli exclusion principle at short
nucleon-$\alpha$ distances, which masks the short-range details of the
nuclear interaction.  On the other hand, the same interaction is not
able to reproduce well the two $P$-wave phase shifts, which are both
too small and too close to each other. 

Recently, the NCSM+RGM approach was applied to a low-energy radiative
capture reaction ${}^7\mathrm{Be} ( p,\gamma)
{}^8\mathrm{B}$~\cite{navratil2011}, and two fusion reactions
${}^3\mathrm{H} ( d,n) {}^4\mathrm{He}$ and ${}^3\mathrm{He} ( d,p)
{}^4\mathrm{He}$~\cite{navratil2012}. In all cases, a SRG evolved
chiral EFT potential at two-body level was considered. It is important
to realize that reaction calculations are extremely sensitive to the
exact positions of thresholds and resonances. In the studies mentioned
above, the authors could utilize the model-dependence on the two-body
SRG flow parameter to reproduce the experimental separation
energies. This will, however, remain a particular challenge for {\em ab
initio} reaction calculations starting from nuclear Hamiltonians
which do not involve free parameters.

\subsection{Neutron rich oxygen and calcium isotopes
\label{subsec:mediumnuclei}}

The oxygen isotopic chain has been extensively studied experimentally
during the last years, see for example
Refs.~\cite{hoffman2009a,kanungo2009}.  The neutron drip
line has been tentatively established at $^{24}$O, but the particle
stability of $^{28}$O is still a matter of a debate since the current
experiments are on the limit of the production cross section. The
isotopes $^{22}$O ($N=14$) and $^{24}$O ($N=16$) exhibit a doubly
magic nature. Their structure is believed to be governed by the
evolution of the $1s_{1/2}$ and $0d_{5/2}$ orbits.  Recent experiments
show that $^{25,26}$O are unbound with respect to
$^{24}$O~\cite{hoffman2008,lunderberg2012}.  On the other hand, the
two-neutron drip line for the fluorine isotopes extends beyond
$^{31}$F. One speculates that the behavior of the neutron drip line
for the oxygen and fluorine isotopic chains arise from a delicate
balance between the proton-neutron and neutron-neutron interactions,
the coupling to the continuum and 3NFs
\cite{otsuka2010,hagen2012a,hagen2012b}.

Another chain of isotopes crucial for theoretical developments is the
calcium chain. It contains several possible closed-shell nuclei beyond
the well established ones, namely $^{40}$Ca and $^{48}$Ca. The $N=32$
sub-shell closure has been established from experiments on
calcium~\cite{huck1985,gade2006}, titanium~\cite{janssens2002}, and
chromium~\cite{prisciandaro2001}. The nucleus $^{52}$Ca has a reduced
value of $2_1^+$ excitation (but more than twice as large as seen in
open-shell calcium isotopes) than that observed in $^{48}$Ca,
suggesting a sub-shell closure.  For $^{54}$Ca ($N=34$), no sub-shell
closure has been seen experimentally in chromium~\cite{marginean2006}
or titanium~\cite{liddick2004,dinca2005}, and there are some doubts
regarding a sub-shell closure in calcium~\cite{rejmund2007}.  The
heaviest calcium isotopes that have been observed are $^{57,58}$Ca
\cite{tarasov2009}, but the masses have been measured only up to
$^{52}$Ca \cite{dilling2012a}.

\begin{figure}[thbp]
  \begin{center}
  \includegraphics[width=0.6\textwidth,clip=]{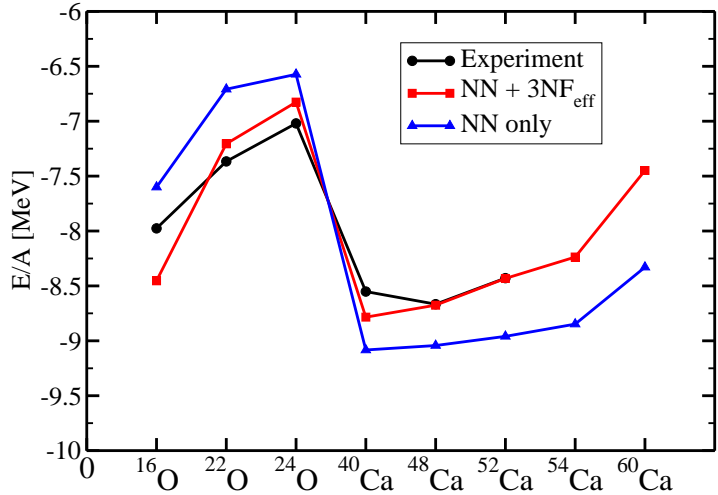}
  \end{center}
  \caption{Binding energy per nucleon for selected oxygen and calcium
    isotopes calculated in coupled cluster theory with NN (triangles)
    and NN+3NF (red squares) forces, compared to experimental data
    (circles).  (Based on Refs.~\cite{hagen2012a,hagen2012b}.)}
\label{fig:beA_oxca}
\end{figure}

\begin{figure}[thbp]
  \begin{center}
\includegraphics[width=0.6\textwidth,clip=]{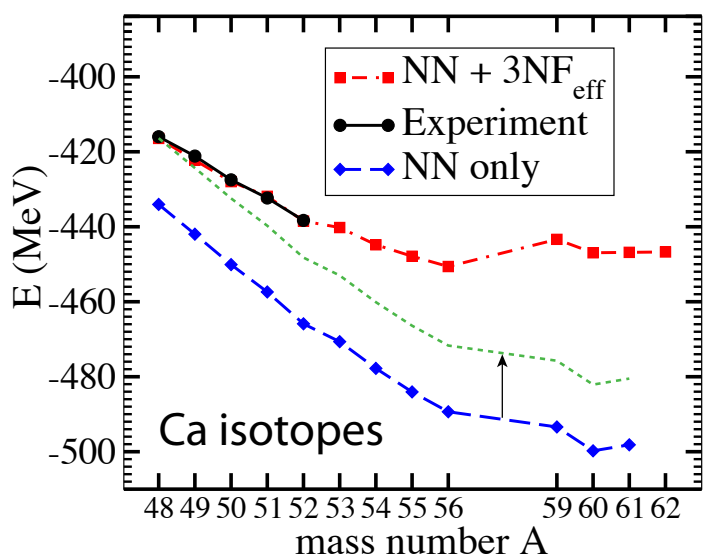}
  \end{center}
  \caption{Binding energies of the neutron-rich calcium isotopes
    calculated in coupled cluster theory with NN (diamonds) and NN+3NF
    (squares) forces, compared to experimental data (circles). The
    dashed line indicates the NN result normalizes at
    $^{48}$Ca. (Based on Ref.~\cite{hagen2012b}.)}
\label{fig:becas}
\end{figure}

As emphasized in Sec.~\ref{sec:formalism}, to describe physics of
nuclei near the drip lines, one needs to properly take into account
(i) the effects of 3NFs, (ii) the presence of open decay channels and
particle continuum, and (iii) many-nucleon correlations.  Recently
coupled-cluster calculations have been carried out for the binding
energies and spectra of the neutron rich oxygen and calcium isotopes
taking for the first time all these effects into account.  The effects
of 3NFs were included effectively in terms of density dependent
corrections to the NN interaction. Those were derived from 3NFs by
summing over the third particle in symmetric nuclear matter. This is
obviously a departure from a rigorous {\em ab initio} approach, but
nevertheless it is a first step towards including effects from both
3NFs and coupling to the particle continuum in coupled-cluster
calculations.  The continuum effects were included by using a Berggren
basis for the relevant partial waves. The binding energies per
particle for selected oxygen and calcium isotopes are displayed in
Fig.~\ref{fig:beA_oxca}. The inclusion of 3NFs significantly improves
agreement with experiment.
In order to investigate the evolution of shell structure and binding
energy systematics in the neutron rich calcium isotopes in more
detail, we show in Fig.~\ref{fig:becas} the total binding energies for
very neutron rich calcium isotopes ranging from $^{48}$Ca to
$^{62}$Ca. Several interesting features are apparent: (i) 3NFs are
important; (ii) a peninsula of weak binding appears for the very
neutron rich calcium isotopes $^{60,61,62}$Ca, and (iii) the results
with NN interactions only, do not yield a flattening of binding energy
at large neutron numbers.  Looking in more detail at the structure of
$^{60-62}$Ca, we find the $J^{\pi}={1/2}_1^+$ ground state of
$^{61}$Ca slightly above the threshold, unbound by about 0.2~MeV with
respect to $^{60}$Ca, and entirely dominated by an $s$ wave.
Furthermore, the ordering of the $gds$ shells is found to be reversed
as compared to the standard shell model ordering. In particular, we
find that the ${5/2}^+$ state appears below the ${9/2}^+$ state in
$^{53,55,61}$Ca.  It is to be noted that the calculations employing
the oscillator basis yield for $^{61}$Ca the level ordering of the
conventional shell model, with a ${9/2}^+$ ground state spin
assignment, thus suggesting strong effects due to the continuum
coupling.  The nuclei $^{61}$Ca and $^{62}$Ca are predicted to be only
weakly unbound with respect to $^{60}$Ca.

\begin{figure}[thbp]
  \begin{center}
 \includegraphics[width=0.6\textwidth,clip=]{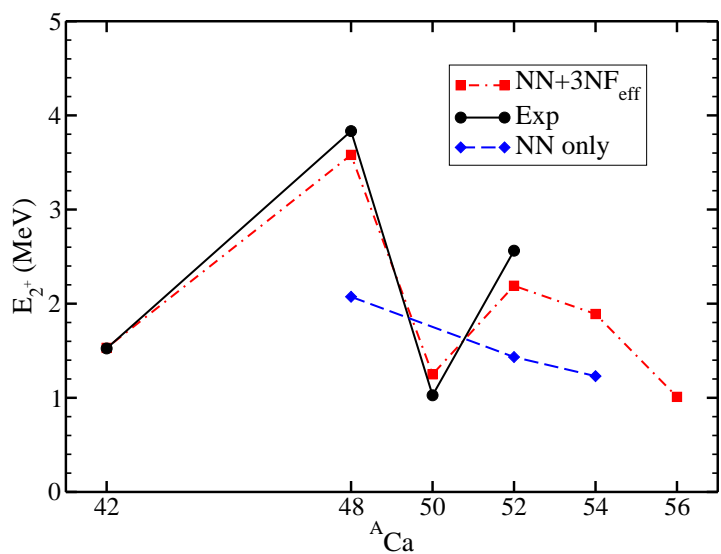}
  \end{center}
  \caption{First $2^+$ excited state energies for the calcium isotopes
    $^{42,48,50,52,54,56}$Ca. Black circles: experimental data; blue
    diamonds: results from nucleon-nucleon interactions; red squares:
    results including the effects of three-nucleon forces.  (Based on
    Ref.~\cite{hagen2012b}.)}
\label{fig:ca2plus}
\end{figure}

Figure~\ref{fig:ca2plus} shows the results for the $2_1^+$ states in
the neutron rich calcium isotopes. It is interesting to note that
coupled-cluster calculations predict only a weak sub-shell closure in
$^{54}$Ca with a $2^+$ state around 1.9 MeV.
These examples clearly point tp the need for 3NFs in neutron-rich
nuclei. Overall, we find that the effects of 3NFs and the scattering
continuum are essential for understanding the evolution of shell
structure towards the drip line. Although more investigations are
clearly needed, our results hint at a situation where odd calcium
isotopes beyond $^{60}$Ca are unbound, while even isotopes can be weakly
bound.

\begin{figure}[htbp]
\centerline{
  \includegraphics[width=0.8\textwidth,clip=]{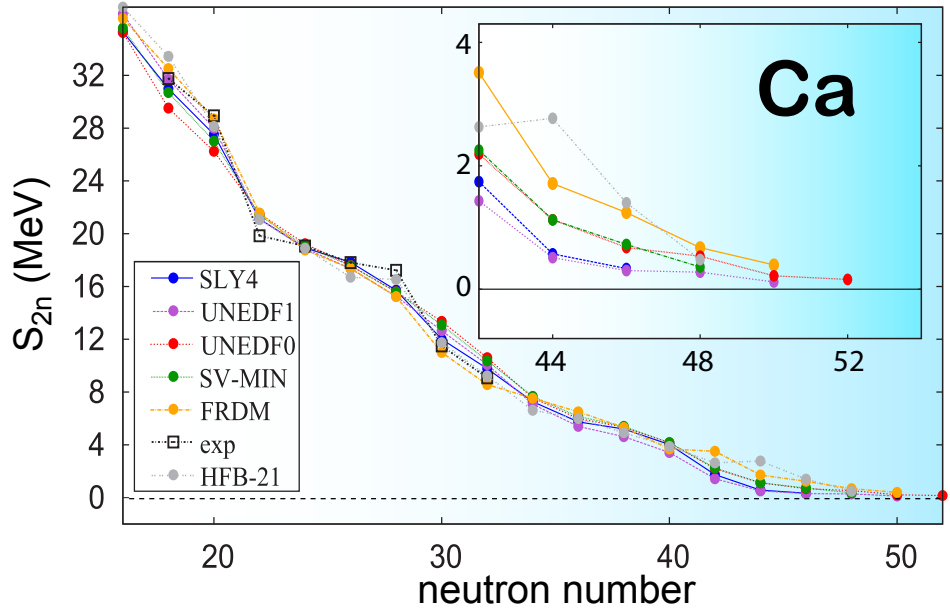} }
\caption{Theoretical extrapolations towards drip lines for the
  two-neutron separation energies $S_{2n}$ for the isotopic chain of
  even-even calcium isotopes using different EDFs (SLy4, SV-min,
  UNEDF0, UNEDF1), and FRDM \cite{moeller1995} and HFB-21
  \cite{goriely2010} mass models (see text for more details). All
  models are consistent with the available experimental data.
  Detailed predictions around $S_{2n}=0$ are illustrated in the inset.
  Drawn by Erik Olsen based on Ref.~\cite{witeknature2012}.}
  \label{fig:s2ncalcium}
\end{figure}

It is interesting to relate the coupled-cluster results for the Ca
isotopes to those based on nuclear DFT.  Figure
\ref{fig:s2ncalcium} shows the two-neutron separation energies for
even-even calcium isotopes computed in Ref.~\cite{witeknature2012}
using SLy4, SV-min, UNEDF0, and UNEDF1 EDFs and obtained in FRDM
\cite{moeller1995} and HFB-21 \cite{goriely2010} mass models.  All
those models predict consistently the neutron drip line around
$^{70}$Ca. Interestingly, in all cases the two-neutron drip lins
($S_{2n}=0$) is approached fairly gradually in all cases; this
indicates that the neutron chemical potential $\lambda_n \approx
-S_{2n}/2$ stays close to, but below, zero for $60\le N \le 70$.
A similar result was obtained in other DFT calculations
\cite{Meng2002,Bhattacharya,Sandulescu,Grasso}. As discussed in
Refs.~\cite{HFBPairing,Fayans2000,Rotival,Zhang2011,LuluLi}, the
persistant appearance of $\lambda_n$ just below the threshold can be
associated with the continuum effect due to pairing. Indeed, the
scattering of neutron pairs into the close-lying non-resonant
continuum gives rise to a stabilization of binding energy and a very
weak dependence of $\lambda_n$ on $N$; hence, extension of the range
of bound nuclei. In terms of HFB quasiparticles, this continuum
coupling manifests itself through increased occupations of low-lying
quasi-particle states, especially those having low orbital angular
momentum \cite{Fayans2000}, as one approaches the threshold. This
results in a gradual increase of contribution from nonresonant
continuum to the ground state of the system.

By looking at canonical HFB states, one can notice the emergence of
bound canonical orbits from the single-neutron continuum. For the
considered case of the drip line Ca nuclei, the canonical neutron
states $s_{1/2}$, $d_{5/2}$, and $d_{3/2}$ appear very close to the
bound $g_{9/2}$ level, which is expected to form the valence shell in
the traditional shell model \cite{Meng2002}.  The fact that
high-$\ell$ and low-$\ell$ orbits bunch up very close to the
$\lambda_n=0$ threshold creates an opportunity for other correlations
to further lower the binding energy.
\begin{figure}[htbp]
\centerline{
  \includegraphics[width=0.7\textwidth]{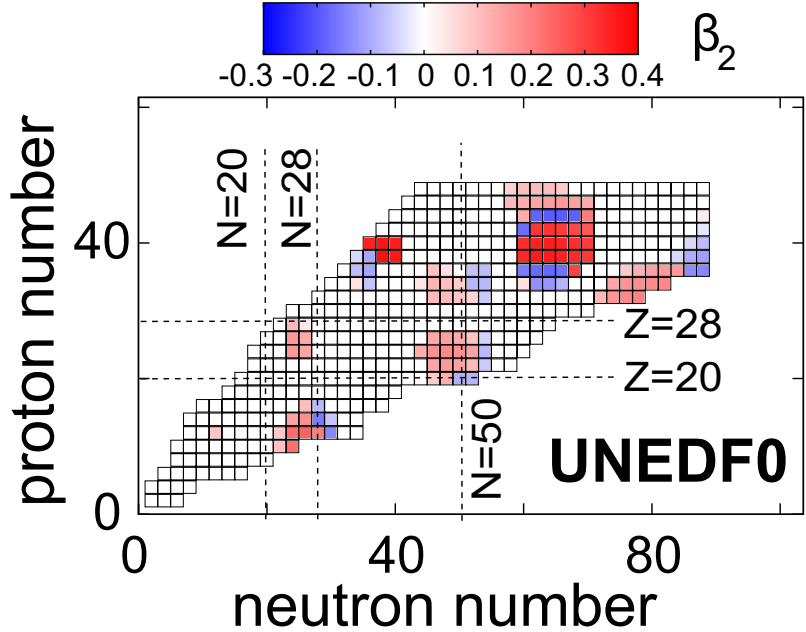} }
\caption{Isoscalar quadrupole deformation $\beta_2$ for $Z<50$
  even-even nuclei predicted in HFB calculations with UNEDF0 energy
  density functional \cite{kortelainen2010}. The neutron-rich Ca
  isotopes with $N\approx 50$ are predicted to be deformed.  (Based on
  Ref.~\cite{witeknature2012}.)}
  \label{fig:defcalcium}
\end{figure}
Figure~\ref{fig:defcalcium} shows the isoscalar (mass) ground-state
quadrupole deformations in $Z<50$ even-even nuclei calculated in the
HFB+UNEDF0 model in Ref.~\cite{witeknature2012}. The appearance of
deformation around $^{70}$Ca is clearly seen. For more discussion on
deformed drip line nuclei, see
Refs.~\cite{Maxmisu,stoitsov2003,witeknature2012,LuluLi,ChenYing2012}.

The binding energy stabilization close to the threshold appears
naturally in the continuum shell model through the collective coupling
of shell model states via the decay channel
\cite{Clustering,Fortschritte}. This coupling, governed by the
anti-Hermitian term in the effective Hamiltonian that represents the
continuum coupling, leads to the formation of the collective aligned
state.  The mechanism responsible for the creation of an aligned
state is similar to the formation mechanism of super-radiant states
\cite{Sokolov,Drozdz}.  In this language, the behavior of drip line
Ca isotopes, the particle stability of $^{6,8}$He and $^{11}$Li, and
the appearance of cluster states near the reaction threshold that
exhausts most of the decay width, are all manifestations of the same
emergent near-threshold phenomenon \cite{Clustering,Fortschritte}.

Clearly, it would be very interesting to see if {\em ab initio} approaches
confirm the trends predicted in nuclear DFT. The fact that the
coupled-cluster calculations of Ref.~\cite{hagen2012b} yield a
$J^\pi=1/2^+$ ground state for $^{61}$Ca, unlike the shell ordering 
in the conventional shell model, is a tantalizing hint that
this is indeed the case.

\section{Conclusions and perspectives} \label{sec:conclusions} 
In this contribution,  we have demonstrated that modern many-body techniques  are nowadays capable of providing a
reliable description of  nuclear properties thanks to  new  conceptual insights  as well as 
algorithmic and computational advances.  For light nuclei, like the chain of lithium and beryllium
isotopes, no-core shell-model calculations 
provide an invaluable tool to link nuclear
structure to  the underlying forces. Similarly, 
coupled cluster theory that  includes  3NFs and coupling to the particle continuum, is capable of 
providing reliable predictions for  heavier nuclei such as oxygen and calcium isotopes. Modern nuclear density functional theory, with optimized energy density functionals, provides an excellent description of heavy nuclei, and it links to 
 {\em ab initio} approaches as well. Recent parametrizations of nuclear energy
density functionals 
provide fairly consistent  predictions when extrapolated to mass regions
where experimental data are not available. 
Close to the limits of stability, the
degrees of freedom represented by resonances, weakly bound states and
the non-resonant continuum need to be accounted for
properly.  The latter has important consequences for the
interpretation of rare isotopes  in terms of a naive shell-model
picture. Our results indicate that this traditional picture may not be relevant
close to  the neutron  drip line.   All of this holds great
promise for a quantitative and predictive modeling of nuclei from a
bottom-up perspective. 

It is important to emphasize that the nuclear many-body problem  is a splendid  example of a multiscale problem, with
length scales spanning many decades. A description of multiscale processes entails
different theoretical methods for different length
scales. First-principle methods are limited to few interacting
nucleons.  With an increasing number of degrees of freedom, DFT-based methods  become the methods of
choice. For even larger systems, a molecular-dynamics-based modeling
is the favored approach. To link these different scales and methods
properly is a great challenge not only to nuclear physics, but to
fields as diverse as material science and life science, see
for example Ref.~\cite{multiscale}.

\begin{figure}[htb]
\centerline{
  \includegraphics[width=0.6\textwidth]{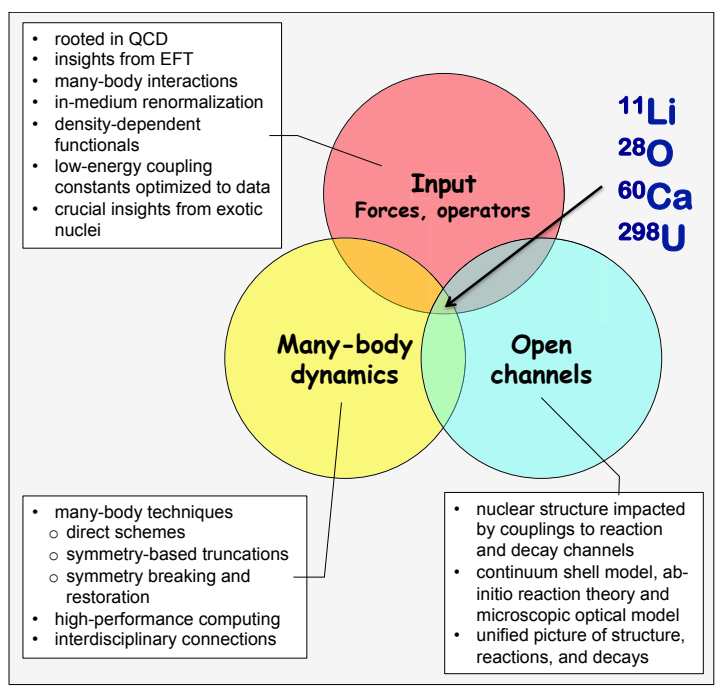} }
\caption{\label{fig:challenges} Challenges of the nuclear many-body problem.
A comprehensive theoretical framework that
would be quantitative, have predictive power, and provide uncertainty quantification 
must meet three stringent requirements: (i) the input must be quantified and of high quality; (ii) many-body dynamics and
correlations must be accounted for; and (iii) the associated formalism
must take care of open-quantum-system aspects of the nucleus.
Only then can we hope to understand rare isotopes, such as $^{11}$Li (two-neutron halo), $^{28}$O (doubly-magic, probably unbound),
$^{60}$Ca (territory for new physics, where {\em ab initio} theory and DFT meet), and $^{298}$U (r-process system, fission recycling participant),
}
\end{figure}
In this work, we have illustrated some challenges and opportunities of the modern nuclear many-body problem, especially in the context of rare isotopes. In particular, we have emphasized the need for a high quality input, the 
importance of  many-body dynamics and the impact of the coupling to open channels, summarized in
Fig.~\ref{fig:challenges}. 
With a fundamental picture of nuclei based on the correct microphysics, we can remove the empiricism inherent today, giving us 
thereby a greater confidence in the science we deliver and the predictions we make.
Guided by unique data on rare isotopes, we are embarking on a comprehensive study of {\it all nuclei}, based on the most accurate knowledge of nuclear interactions, the most reliable theoretical approaches, and  massive use of new computer 
hardware and advanced numerical algorithms. The prospects are excellent.

\bigskip
\bigskip

\noindent {\bf Acknowledgments}

We are much indebted to Carlo Barbieri, Bruce Barrett, Scott Bogner, Alex Brown,
David Dean, Jacek Dobaczewski, Thomas Duguet, Dick Furnstahl, Alexandra Gade, Mihai
Horoi, Gustav Jansen, Robert Janssens, \O yvind Jensen, Bj{\"o}rn Jonson, Simen Kvaal, 
Augusto Macchiavelli, Pieter Maris,
Nicolas Michel, Petr Navrat\'il, Thomas Nilsson, Filomena Nunes,
Takaharu Otsuka, George Papadimitriou, Thomas Papenbrock,  Lucas Platter, Marek
P{\l}oszajczak, Sofia Quaglioni, Robert Roth, Achim Schwenk, Brad Sherill, Olivier Sorlin,
Artemisia Spyrou, Michael Thoennessen, Jeff Tostevin, Koshiroh Tsukiyama, and James Vary for the many 
stimulating discussions
on nuclear physics, many-body physics, shell-model and coupled-cluster calculations,
nuclear density functional theory, and open quantum systems.  The
research leading to these results has received funding from the European
Research Council under the European Community's Seventh Framework
Programme (FP7/2007-2013) / ERC grant agreement no.~240603. 
This work was supported by the U.S.~Department of Energy under Contract Nos.~DE-AC05-00OR22725
(Oak Ridge National Laboratory, managed by UT--Battelle, LLC),
DE-FG02-96ER40963 (University of Tennessee), DE-FG05-87ER40361 (Joint
Institute for Heavy Ion Research), DE-FC02-09ER41583 (UNEDF SciDAC
Collaboration), DE-FG02-04ER41338, the US NSF under grant
PHY-0854912, and by the Swedish Research Council (dnr.~2007-4078). Computational resources were provided by the Oak Ridge
Leadership Class Computing Facility, the National Energy Research
Scientific Computing Facility, and the Research Council of Norway via
the Notur project (Supercomputing grant NN2977K).  This work was
supported by the Office of Nuclear Physics, U.S.~Department of Energy
(Oak Ridge National Laboratory).

\bigskip

\bibliographystyle{iop} 
\bibliography{NobelTheory}

\end{document}